\documentclass[12pt]{scrartcl}

\usepackage[ruled,vlined,linesnumbered]{algorithm2e}
\usepackage[centertags]{amsmath}
\usepackage{amssymb}
\usepackage[usenames,dvipsnames,table]{xcolor}
\usepackage{hyperref}
\hypersetup{
    bookmarks=false,      
    pdfstartview={FitH},  
    colorlinks=true,      
    linkcolor=black,      
    citecolor=blue,       
    filecolor=blue,       
    urlcolor=blue         
}

\usepackage{varioref}
\usepackage{cite}
\usepackage{longtable}
\usepackage{pifont}
\usepackage{paralist}

\usepackage{mathtools}
\usepackage{fancyvrb}
\usepackage{dsfont}

\usepackage{subfigure}

\usepackage{tikz}
\usetikzlibrary{shapes,arrows}

\definecolor{tableshade1}{HTML}{F1F5FA} 
\definecolor{tableshade2}{HTML}{ECF3FE} 

\tikzstyle{decision} = [diamond, draw, fill=blue!20, 
    text width=4.5em, text badly centered, node distance=3cm, inner sep=0pt]
\tikzstyle{block} = [rectangle, draw, fill=blue!20, 
    text width=5em, text centered, rounded corners, minimum height=4em]
\tikzstyle{line} = [draw, -latex']
\tikzstyle{cloud} = [draw, ellipse,fill=red!20, node distance=3cm,
    minimum height=2em]

\newcommand{\Red}[1]{\textcolor{red}{#1}} 
\newcommand{\cpp}{\texttt{C++}}
\newcommand{\bs}[1]{\ensuremath{\boldsymbol{#1}}}
\newcommand{\bsn}[2]{\ensuremath{\boldsymbol{#1^{(#2)}}}}

\newcommand{\SU}[1]{\ensuremath{\mathrm{SU}(#1)}}
\newcommand{\U}[1]{\ensuremath{\mathrm{U}(#1)}}
\newcommand{\code}[1]{{\small{\texttt{#1}}}}
\newcommand{\gid}[2]{\ensuremath{\mathfrak{G}({#1},{#2})}}
\newcommand{\gap}{{\small{\texttt{GAP}}}}
\newcommand{\python}{\code{Python}}
\newcommand{\gapid}{{\small{\texttt{GAPID}}}}
\newcommand{\vev}{{\small{\texttt{vev}}}}
\newcommand{\vevs}{{\small{\texttt{vevs}}}}
\newcommand{\UPMNS}{\ensuremath{U_\mathrm{PMNS}}}
\newcommand{\UCKM}{\ensuremath{U_\mathrm{CKM}}}
\newcommand{\UHPS}{\ensuremath{U_\mathrm{HPS}}}
\newcommand{\UL}{\ensuremath{U_\mathrm{L}}}
\newcommand{\DL}{\ensuremath{D_\mathrm{L}}}
\newcommand{\ULD}{\ensuremath{U_\mathrm{L}^\dagger}}

\newcommand{\m}{\text{-}}
\newcommand{\p}{\text{\phantom{-}}}

\labelformat{section}{Section #1} 
\labelformat{appendix}{Section #1} 
\labelformat{subsection}{Section #1} 
\labelformat{subsubsection}{Section #1} 
\labelformat{equation}{Eq.~(#1)} 
\labelformat{figure}{Fig.~#1} 
\labelformat{table}{Tab.~#1} 
\labelformat{subfigure}{Fig.~\thefigure(\alph{subfigure})}

\setkomafont{subsection}{\normalcolor\bfseries}
\addtokomafont{caption}{\footnotesize}



\begin{document}

\titlehead{\hfill LPSC-10210}
\title{Tribimaximal Mixing From Small Groups}
\author{Krishna Mohan Parattu\footnote{krishna@iucaa.ernet.in}\\
{\normalsize \it Inter-University Centre for Astronomy and Astrophysics}\\[-1ex]
{\normalsize \it Ganeshkhind, Pune 411007, India}\\
\and
Ak\i{}n Wingerter\footnote{akin@lpsc.in2p3.fr}\\
{\normalsize \it Laboratoire de Physique Subatomique et de Cosmologie}\\[-1ex]
{\normalsize \it UJF Grenoble 1, CNRS/IN2P3, INPG}\\[-1ex]
{\normalsize \it 53 Avenue des Martyrs, F-38026 Grenoble, France}\\
}
\date{}
\dedication{{\bfseries Abstract}\\[2ex]
\begin{minipage}[b]{0.9\linewidth} 
\small Current experimental data on the neutrino parameters is in good
agreement with tribimaximal mixing and may indicate the presence of an
underlying family symmetry. For 76 flavor groups, we perform a
systematic scan for models: The particle content is that of the
Standard Model plus up to three flavon fields, and the effective
Lagrangian contains all terms of mass dimension $\leq6$. We find that
44 groups can accommodate models that are consistent with experiment
at $3\sigma$, and 38 groups can have models that are tribimaximal. For
one particular group, we look at correlations between the mixing
angles and make a prediction for $\theta_{13}$ that will be testable
in the near future. We present the details of a model with
$\theta_{12}=33.9^\circ$, $\theta_{23}=40.9^\circ$,
$\theta_{13}=5.1^\circ$ to show that the recent tentative hints of a
non-zero $\theta_{13}$ can easily be accommodated. The smallest group
for which we find tribimaximal mixing is $T_7$. We argue that $T_7$
and $T_{13}$ are as suited to produce tribimaximal mixing as $A_4$ and
should therefore be considered on equal footing. In the appendices, we
present some new mathematical methods and results that may prove
useful for future model building efforts.
\end{minipage} }

\maketitle[0]
\newpage
\tableofcontents 

\clearpage
\newpage
\section{Introduction}

Neutrino physics is a fast developing field. The past decade has seen the discovery of neutrino masses \cite{Fukuda:1998mi,Ahmad:2002jz} and ever improving measurements of the neutrino mixing matrix \UPMNS{} \cite{Pontecorvo:1957cp,Maki:1962mu}. Our growing knowledge of the neutrino parameters \cite{Schwetz:2008er,GonzalezGarcia:2010er} has almost raised more questions than it answered: Why are neutrinos so light? Why are two of the mixing angles large and one vanishingly small? Why is \UPMNS{} so different from \UCKM{} \cite{Kobayashi:1973fv}? These are some of the questions that any model for the neutrino sector needs to address.

Experimental data suggests that the mixing angles are in good agreement with tribimaximal mixing (TBM) \cite{Harrison:2002er,Harrison:2002kp}. The very form of the Harrison-Perkins-Scott matrix $\UHPS{}$ is suggestive of an underlying family symmetry between the three generations of leptons. In the past years, much effort has been vested in finding a family symmetry that would naturally lead to tribimaximal mixing, and to that end, some twenty odd groups have been the subject of model building efforts (see refs.~\cite{Altarelli:2010gt,Ishimori:2010au,Ludl:2010bj} and references therein).

\medskip

It has been argued that $A_4$ is particularly relevant for producing tribimaximal mixing \cite{Ma:2001dn,Ma:2004pt,Babu:2005se,Altarelli:2005yx}, and by the number of publications (see e.g.~Tab.~2 in ref.~\cite{Altarelli:2010gt}) it is certainly the most popular discrete symmetry used for model building. That is why we start out by following down the same path to construct all $A_4\times \mathbb{Z}_3$ models with up to three flavon fields where the lepton doublet $L$ transforms as a triplet. We find 22,932 inequivalent models, of which 4,481 (19.5\%) give mixing angles that are consistent with experiment at $3\sigma$, and 4,233 (18.5\%) that are tribimaximal. Restricting $\theta_{12}$ and $\theta_{23}$ to their respective $3\sigma$ intervals, we obtain an interesting prediction for $\theta_{13}$ whose value is currently not known with very high precision: The by far most likely value is $\theta_{13}=0^\circ$, and there are extremely few models for $0^\circ \lneqq \theta_{13} \lesssim 12^\circ$. We also present a model where all three mixing angles $\theta_{12} \simeq 34^\circ$, $\theta_{23} \simeq 41^\circ$ and $\theta_{13} \simeq 5^\circ$ lie in their respective $1\sigma$ intervals to show that it is possible to accommodate the recent tentative hints of a non-zero $\theta_{13}$ \cite{GonzalezGarcia:2010er,Mezzetto:2010zi,Fogli:2010zz}.

\medskip

Next we explore whether $A_4$ is really special or we are looking for tribimaximal models ``under the lamppost''. There are 1,048 groups with less than or equal to 100 elements, and 90 of them have a 3-dimensional irreducible representation. For 76 groups, we construct all models with up to three flavon fields where the lepton doublet $L$ transforms in a 3-dimensional irreducible representation. For the remaining 14 groups, a systematic scan would simply take too long. We find 44 groups (58\%) that can accommodate models which are consistent with experiment at the $3\sigma$ level, and 38 groups (50\%) that can produce tribimaximal mixing. The smallest group for which we find tribimaximal mixing is $T_7$, and the group with the largest fraction of tribimaximal models is $T_{13}$. Incidentally, for $T_{13}$ (and the other metacyclic groups) the set of tribimaximal models and the set of $3\sigma$ models are almost identical, and this may be pointing towards a profound connection between $T_{13}$ and tribimaximal mixing that is more pronounced as compared to $A_4$. For a recent publication that uses $T_{13}$ for model building, see ref.~\cite{Kajiyama:2010sb}. 

\medskip

For our analysis, the computer algebra program \gap{} \cite{GAP4} played a central role. We used \gap{} to obtain the character table, the dimension of the conjugacy classes and the explicit form of the representation matrices for the 76 groups that we considered in this publication. In contrast to e.g.~solving renormalization group equations, the use of computers for algebraic and group theoretic operations is not widespread (a notable exception is ref.~\cite{Ludl:2009ft}). We strongly advocate the use of the SmallGroups Library \cite{GAP4:smallgroups} which collects in one place and provides easy access to all finite groups of order at most 2,000 (except 1,024). 

\medskip

In the appendices we present some new developments and mathematical background information relevant for model building with discrete symmetries. 

In \ref{app:90groups}, we list the 90 groups of order less than or equal to 100 that have a 3-dimensional irreducible representation. For each group, we indicate whether it is a subset of \U{3}, \U{2} or $\U{2}\times \U{1}$, and at the same time check whether it contains $A_4$ as a subgroup. Due to its length, the full list of the 1,048 groups of order at most 100 is presented in a separate file \cite{aw:2010:symmetries}. 

In \ref{sec:allgroups} we give the full details on how we generated the 1,048 groups and compiled the tables in \ref{app:90groups} and ref.~\cite{aw:2010:symmetries}. We elaborate on some disagreement that we have with the existing literature.

In \ref{sec:breaking_the_family_symmetry} we show how to find the vacuum expectation values that break a given group to any one of its subgroups. Unfortunately, finding all possible symmetry breaking patterns does not allow us to classify the models, since different \vevs{} inducing the same symmetry breaking chain may lead to different mixing angles.

In \ref{app:clebsch-gordan-coefficients}, we discuss an algorithm due to van den Broek and Cornwell \cite{vandenbroek:1978aa} for calculating the Clebsch-Gordan coefficients for any finite group. This allows us to construct the group invariants, or more generally, contract the family indices in the Lagrangian without referring to heuristic constructions as is common practice in the current literature.

Finally, in \ref{app:elements_of_finite_group_theory} we outline some of the most important concepts and theorems from the theory of groups that pertain to the present publication.

\section{Experimental Constraints}
\label{sec:experimental_results}

The leptonic mixing matrix \UPMNS{} is generally parametrized by three angles, $\theta_{12}$, $\theta_{23}$, $\theta_{13}$, and one Dirac phase $\delta$~\cite{Nakamura2010}. If the neutrinos are Majorana particles, there are two extra phases $\phi_1$ and $\phi_2$ that do not affect neutrino oscillation phenomena\cite{Bilenky:1980cx} and are likely to remain unconstrained in the near future. In this paper, we use the standard parametrization \cite{Nakamura2010} of \UPMNS{} except for the definition of the Majorana phases, where we follow ref.~\cite{Plentinger:2006nb}:

\medskip
\begin{equation}
\small
\UPMNS= \left(\begin{array}{lll}

c_{12}c_{13}       & s_{12}c_{13}             & s_{13} e^{\m i\delta} \\
\m s_{12}c_{23}-c_{12}s_{13}s_{23}e^{i\delta}   & c_{12}c_{23}-s_{12}s_{13}s_{23}e^{i\delta}  & c_{13}s_{23}\\
s_{12}s_{23}-c_{12}s_{13}c_{23}e^{i\delta}    & \m c_{12}s_{23}-s_{12}s_{13}c_{23}e^{i\delta} & c_{13}c_{23}

\end{array}\right)\cdot\textrm{diag}(e^{i \phi_{1}},e^{i \phi_{2}},1)
\end{equation}

\medskip

For comparing our results from \ref{sec:analysis} to experiment, we used refs.~\cite{GonzalezGarcia:2010er,Schwetz:2008er,Mezzetto:2010zi}. In \ref{tab:experimental_data}, we summarize the relevant information for the reader's convenience.

\begin{table}[h]
 \centering
\begin{tabular}{|c|c|c|c|}\hline
 Parameter      &  Mean Value          &  1$\sigma$ range         &   3$\sigma$ range                     \\\hline \hline
 $\theta_{12}$  &  $34.4^\circ$        &  $33.4^\circ-35.4^\circ$ &   $31.5^\circ-37.6^\circ$             \\\hline
 $\theta_{23}$  &  $42.8^\circ$        &  $39.9^\circ-47.5^\circ$ &   $35.5^\circ-53.5^\circ$             \\\hline
 $\theta_{13}$  &  $5.6^\circ$         &  $2.9^\circ-8.6^\circ$   &   \phantom{1.2}$0^\circ-12.5^\circ$    \\\hline
\end{tabular}
\caption{\footnotesize The leptonic mixing angles from the global fit to data from ref.~\cite{GonzalezGarcia:2010er} (first table, left column).}
\label{tab:experimental_data}
\end{table}

The solar and atmospheric neutrino mixing angles, $\theta_{12}$ and $\theta_{23}$, are relatively well-determined. $\theta_{13}$, on the other hand, effectively only has an upper bound. The experimental data is consistent with $\theta_{13}$ being zero, as e.g.~in exact tribimaximal mixing. If $\theta_{13}=0^\circ$, the Dirac phase loses physical significance. Currently, there are possible hints for a non-zero $\theta_{13}$ \cite{GonzalezGarcia:2010er,Mezzetto:2010zi,Fogli:2010zz}. A new generation of neutrino experiments will probe $\sin^{2}\theta_{13}$ down to about $10^{-2}$~\cite{Mezzetto:2010zi}.

\section{A Paradigm: $\bs{A_4\times C_3}$ Family Symmetry with Three Flavon Fields}
\label{sec:altarelli-feruglio-model}

To illustrate our general approach, we will choose $A_4\times C_3$ as the family symmetry and reproduce the results of the now classic paper by Altarelli and Feruglio \cite{Altarelli:2005yx}. Here and in the following we will use the alternate notation $C_n$ for $\mathbb{Z}_n$. Note that we could have taken any of the 439,820 models that we will be constructing in \ref{sec:sym_lag_mix}, but considering a model that is already well-known has the advantage of a clearer presentation of our methodology by stressing the differences to other approaches.

\medskip

The following lines of \gap{} code give us information on the group $A_4\times C_3$:

\begin{Verbatim}[fontsize=\scriptsize,numbers=left,xleftmargin=20pt,formatcom=\color{gray}]
group := SmallGroup(36,11);;
Display(StructureDescription(group));
chartab := Irr(group);;
Display(chartab);
SizesConjugacyClasses(CharacterTable(group));
LoadPackage("repsn");;
for i in [1..Size(chartab)] do
  rep := IrreducibleAffordingRepresentation(chartab[i]);
  for el in Elements(group) do
    Display(el^rep);
  od;
od;
\end{Verbatim} 
\label{verb:gap_group_properties}

These lines can be entered directly at the \gap{} prompt or saved in a file and executed automatically as explained later. Line 1 defines the group in terms of its \gapid{} (see \ref{sec:generate_groups}). Lines 4 and 5 display the character table and the dimensions of the conjugacy classes, respectively. Finally, lines 6-12 give the explicit form of the matrices for all elements and for all representations of the group.

\medskip

The first column of the character table gives the dimensions of the representations. We follow the common practice of denoting the representations by their dimensions and using primes or numbers to distinguish different representations of the same dimension:
\begin{equation}
\bs{1} , \quad  \bs{1'} , \quad  \bs{1''} , \quad  \bs{1'''} , \quad  \bsn{1}{4} , \quad  \bsn{1}{5} , \quad  \bsn{1}{6} , \quad  \bsn{1}{7} , \quad  \bsn{1}{8} , \quad  \bs{3} , \quad  \bs{3'} , \quad  \bs{3''}
\end{equation}
Note that we deviate from the notation of ref.~\cite{Altarelli:2005yx} where the transformation properties of the representation under the factor subgroups are indicated, e.g.~$\bs{3}\otimes\omega$, where $\omega$ is the primitive third root of unity. The reason why we choose another notation is that we would like to deal with all groups on equal footing. It is easy to establish the connection between the two notations by comparing the representation matrices of $A_4$ and $A_4\times C_3$ e.g.~for the $\bs{3}$: The first, third and fourth generator of $A_4\times C_3$ are identical to the three generators\footnote{In the presentation we have chosen, $A_4$ is given by three generators. One can get more information on $A_4$ by running the \gap{} script \vpageref{verb:gap_group_properties} with the \gapid{} [12,3].} of $A_4$ and the second generator generates $C_3$. We can now easily identify $\bs{3}\sim\bs{3}\otimes1$, $\bs{3'}\sim\bs{3}\otimes\omega$, $\bs{3''}\sim\bs{3}\otimes\omega^2$. The other cases are handled in a completely analogous way (see \ref{tab:particle_content} for the complete list). Strictly speaking, though, making this connection is not necessary.

\medskip

\begin{table}
 \centering
 \renewcommand{\arraystretch}{1.2}
\begin{tabular}{|c||c|c|l|l||l|}\hline
 Field        &  $\SU{2}_L\times\U{1}_Y$   & $\U{1}_R$ &  $A_4$            &  $C_3$          &  $A_4\times C_3$               \\\hline \hline
 $L$          &  $(\bs{2},\m1)$            & 1         &  $\bs{3}$         &  $\omega$       &  $\phantom{A_4}\bs{3'}$        \\\hline
 $e$          &  $(\bs{1},\p2)$            & 1         &  $\bs{1}$         &  $\omega^{2}$   &  $\phantom{A_4}\bs{1'}$        \\\hline
 $\mu$        &  $(\bs{1},\p2)$            & 1         &  $\bs{1''}$       &  $\omega^{2}$   &  $\phantom{A_4}\bsn{1}{8}$     \\\hline
 $\tau$       &  $(\bs{1},\p2)$            & 1         &  $\bs{1'}$        &  $\omega^{2}$   &  $\phantom{A_4}\bsn{1}{5}$     \\\hline
 $h_u$        &  $(\bs{2},\p1)$            & 0         &  $\bs{1}$         &  $1$            &  $\phantom{A_4}\bs{1}$         \\\hline
 $h_d$        &  $(\bs{2},\m1)$            & 0         &  $\bs{1}$         &  $1$            &  $\phantom{A_4}\bs{1}$         \\\hline
 $\varphi_T$  &  $(\bs{1},\p0)$            & 0         &  $\bs{3}$         &  $1$            &  $\phantom{A_4}\bs{3}$         \\\hline
 $\varphi_S$  &  $(\bs{1},\p0)$            & 0         &  $\bs{3}$         &  $\omega$       &  $\phantom{A_4}\bs{3'}$         \\\hline
 $\xi$        &  $(\bs{1},\p0)$            & 0         &  $\bs{1}$         &  $\omega$       &  $\phantom{A_4}\bs{1''}$        \\\hline
\end{tabular}
\setcapindent{0em}
\caption{Particle content and charges for the model given in ref.~\cite{Altarelli:2005yx}. The last column gives the family symmetry charges in our notation. In ref.~\cite{Altarelli:2005yx}, there is an evident typo in the charge assignments to $\mu$ and $\tau$ in Section 4 as compared to Section 3 in the same publication.}
\label{tab:particle_content}
\end{table}

The particle content of the model is given in \ref{tab:particle_content}. In the following we list the terms that \begin{inparaenum}[(i)] \item are invariant under the Standard Model gauge symmetry, the $R$-symmetry and the family symmetry, \item contain exactly 2 leptons, \item have mass dimension smaller than or equal to 6, \item are at most linear in the flavon \vevs{}: \end{inparaenum}
\begin{equation}
L \, L \, h_u \, h_u \, \varphi_S + L \, L \, h_u \, h_u \, \xi + L \, e \, h_d \, \varphi_T + L  \,\mu \, h_d  \,\varphi_T + L \, \tau \, h_d \, \varphi_T
\label{eq:invariant_lagrangian}
\end{equation}

\medskip

To check invariance under the family symmetry we need the decomposition of tensor products into irreducible representations (see e.g.~ref.~\cite{Jones:1990ti}) that is readily obtained from the character table and the dimensions of the conjugacy classes. E.g.~for the first term in \ref{eq:invariant_lagrangian} we have:
\begin{equation}
\bs{3'} \otimes \bs{3'} \otimes \bs{1} \otimes \bs{1} \otimes \bs{3'} = \left( \bs{1'}  +  \bsn{1}{5}  +  \bsn{1}{8}  +  2\times \bs{3''} \right) \otimes \bs{3'} = 2\times\bs{1}  +  2\times\bs{1'''}  +  2\times\bsn{1}{4}  +  7 \times \bs{3}
\end{equation}
The tensor product contains 2 singlets and thus there are 2 ways to contract the family indices to obtain invariant combinations. To do this, however, we need to know the Clebsch-Gordan coefficients for $A_4\times C_3$, and to our surprise, the general method for the calculation of Clebsch-Gordan coefficients for any finite symmetry group is not well-known. That is why we have dedicated \ref{app:clebsch-gordan-coefficients} to discuss an algorithm \cite{vandenbroek:1978aa} for the calculation of Clebsch-Gordan coefficients for finite groups. The first term in \ref{eq:invariant_lagrangian} after contracting the family indices becomes:
\begin{align}
\frac{1}{\sqrt{3}} L_{2} \, L_{3} \, h_u \, h_u \, \varphi_{S,1}
+ \frac{1}{\sqrt{3}} L_{3} \, L_{1} \, h_u \, h_u \, \varphi_{S,2}
&+ \frac{1}{\sqrt{3}} L_{1} \, L_{2} \, h_u \, h_u \, \varphi_{S,3}
+ \frac{1}{\sqrt{3}} L_{1} \, L_{1} \, h_u \, h_u \, \xi \notag\\
&+ \frac{1}{\sqrt{3}} L_{2} \, L_{2} \, h_u \, h_u \, \xi 
+ \frac{1}{\sqrt{3}} L_{3} \, L_{3} \, h_u \, h_u \, \xi
\label{eq:invariant_lagrangian_after_A4_contractions}
\end{align}

After contracting the \SU{2} indices and substituting the {\it vevs}
\begin{equation}
\small
\langle h_u^{(1)}\rangle=\langle h_d^{(2)}\rangle=0, \enspace \langle h_u^{(2)}\rangle=v_u, \enspace \langle h_d^{(1)}\rangle=v_d, \enspace \langle\varphi_T\rangle = (v_T, v_T, v_T), \enspace \langle\varphi_S\rangle = (v_S, 0, 0), \enspace \langle\xi\rangle = v_\xi,
\label{eq:altarelli_vevs}
\end{equation}

\ref{eq:invariant_lagrangian_after_A4_contractions} reads:
\begin{equation}
\frac{1}{\sqrt{3}} L_{2}^{(1)} \, L_{3}^{(1)}   \, v_u \, v_u  \, v_S
+ \frac{1}{\sqrt{3}} L_{1}^{(1)} \, L_{1}^{(1)} \, v_u \, v_u \, \xi
+ \frac{1}{\sqrt{3}} L_{2}^{(1)} \, L_{2}^{(1)} \, v_u \, v_u \, \xi
+ \frac{1}{\sqrt{3}} L_{3}^{(1)} \, L_{3}^{(1)} \, v_u \, v_u \, \xi
\end{equation}

Following these steps for all the terms in \ref{eq:invariant_lagrangian} yields the mass matrices for the charged leptons and the neutrinos:
\begin{equation}
M_{\ell^+} = \bordermatrix{
            &       e                      &   \mu                    &    \tau                  \cr
L_1^{(2)}   &       \m\frac{1}{\sqrt{3}}   &   \m\frac{1}{\sqrt{3}}   &    \m\frac{1}{\sqrt{3}}  \cr       
L_2^{(2)}   &       \m\frac{1}{\sqrt{3}}   &   \p\frac{1}{2\sqrt{3}}  &    \p\frac{1}{2\sqrt{3}} \cr
L_3^{(2)}   &       \m\frac{1}{\sqrt{3}}   &   \p\frac{1}{2\sqrt{3}}  &    \p\frac{1}{2\sqrt{3}} \cr
},\qquad
M_{\nu} = \bordermatrix{
            &        L_1^{(1)}              &   L_2^{(1)}             &    L_3^{(1)}              \cr
L_1^{(1)}   &        \frac{1}{\sqrt{3}}     &   \,\,0                 &    \,\,0                  \cr     
L_2^{(1)}   &        \,\,0                  &   \frac{1}{\sqrt{3}}    &    \frac{1}{2\sqrt{3}}    \cr     
L_3^{(1)}   &        \,\,0                  &   \frac{1}{2\sqrt{3}}   &    \frac{1}{\sqrt{3}}     \cr  
}
\label{eq:mass_matrices}
\end{equation}

The singular value decomposition (here in the special case where the number of rows is equal to the number of columns)
\begin{equation}
\hat{M}_{\ell^+} = D_L M_{\ell^+} D_R^\dagger, \qquad \hat{M}_{\nu} = U_L M_{\nu} U_R^\dagger,
\end{equation}
allows us to express the mass matrices as a product of a unitary matrix, a diagonal matrix with non-negative real numbers on the diagonal, and another unitary matrix where
\begin{equation}
\DL = \left(\begin{smallmatrix}
\m0.5774 + i\, 0.0000  &    \m0.5774 + i\, 0.0000   &   \m0.5774 + i\, 0.0000     \\ 
\p0.5738 - i\, 0.0636  &    \m0.2319 + i\, 0.5287   &   \m0.3420 - i\, 0.4652     \\ 
\p0.5731 - i\, 0.0702  &    \m0.3474 - i\, 0.4612   &   \m0.2257 + i\, 0.5314     \\ 
\end{smallmatrix}\right), \qquad
\UL = \left(\begin{smallmatrix}
\p0.0000    &   \m0.7071    &   \m0.7071       \\ 
\m1.0000    &   \p0.0000    &   \p0.0000      \\  
\p0.0000    &   \p0.7071    &   \m0.7071       \\ 
\end{smallmatrix}\right).
\end{equation}

The neutrino mixing matrix is by definition
\begin{equation}
\UPMNS = \DL \ULD = \left(\begin{matrix}
 \p0.8165 + i\, 0.0000  &    \p0.5774 + i\, 0.0000  &    \p0.0000 + i\, 0.0000      \\
 \p0.4058 - i\, 0.0449  &    \m0.5738 + i\, 0.0636  &    \p0.0778 + i\, 0.7028      \\
 \p0.4052 - i\, 0.0497  &    \m0.5731 + i\, 0.0702  &    \m0.0860 - i\, 0.7019      \\
\end{matrix}\right).
\label{eq:UPMNS-altarelli-feruglio}
\end{equation}
In \ref{sec:euler_angles_and_recognizing_tbm}, we discuss how to extract the mixing angles and phases from the most general form for \UPMNS{}. In the present case, we obtain
\begin{equation}
\theta_{12} = 0.00, \quad    \theta_{23} = 35.26, \quad   \theta_{13} = 45.00, \quad   \delta = \m90.00,
\end{equation}
which is tribimaximal mixing. 

\medskip

Several remarks are in order.
\begin{inparaenum}[(i)] 
\item At first glance, \UPMNS{} in \ref{eq:UPMNS-altarelli-feruglio} bears little resemblance to the Harrison-Perkins-Scott matrix \UHPS{}. Note, though, that rephasing the fields, \ref{eq:UPMNS-altarelli-feruglio} can easily be brought to \UHPS{} form. In other cases, there may also be ordering ambiguities (see \ref{sec:euler_angles_and_recognizing_tbm} for more details).
\item In contrast to ref.~\cite{Altarelli:2005yx} where the matrices for the generators in the 3-dimensional representation were wisely chosen so that $M_{\ell^+}$ is diagonal, our choice for the generators leads to a non-diagonal charged lepton mass matrix, see \ref{eq:mass_matrices}. We have checked that after changing to a basis where $M_{\ell^+}$ is diagonal (which corresponds to a redefinition of the charged lepton fields), our expressions for the Lagrangian, the mass matrices and \UPMNS{} coincide with those in ref.~\cite{Altarelli:2005yx}.
\item The same change of basis maps our \vevs{} in \ref{eq:altarelli_vevs} to those of ref.~\cite{Altarelli:2005yx}.
\item The reason why our intermediate results do not coincide with those in ref.~\cite{Altarelli:2005yx} is that we started out with a different choice of generators. Our generators $T_1$, $T_2$, $T_3$ of $A_4$ are connected to the latter ones by $S\mapsto T_1 T_2^{-1} T_1$ and $T\mapsto T_2$.
\item For later reference, we summarize in \vref{tab:altferug-symbreak} the symmetry breaking patterns for the model at hand. In \vref{sec:breaking_the_family_symmetry} we will discuss in detail how to find all inequivalent \vevs{} that break to different subgroups of a given symmetry.
\end{inparaenum}

\medskip

\begin{table}[h!]
\begin{center}
\renewcommand{\arraystretch}{1.2}
\begin{tabular}{|l|l||l|l||l|l|}
\hline
\multicolumn{2}{|c||}{\textsc{1 \vev{}}} & \multicolumn{2}{c||}{\textsc{2 \vevs{}}} & \multicolumn{2}{c|}{\textsc{3 \vevs{}}}\\
\hline
\hline
$\varphi_T$      & $\mathfrak{h}_{22}=C_3\times C_3$              & $\varphi_T$, $\xi$       & $\mathfrak{h}_{10}=C_3$         &  &  \\
\cline{1-4}
$\varphi_S$      & $\mathfrak{h}_{4\phantom{2}}=C_2$              & $\varphi_S$, $\xi$   & $\mathfrak{h}_{4\phantom{2}}=C_2$ & $\varphi_T$, $\varphi_S$, $\xi$ & $\mathfrak{h}_{1}=1$ \\
\cline{1-4}
$\xi$            & $\mathfrak{h}_{28}=A_4$                        & $\varphi_T$, $\varphi_S$ & $\mathfrak{h}_{1\phantom{4}}=1$ &   & \\
\hline
\end{tabular}
\end{center}
\setcapindent{0em}
\caption{The subgroups to which $A_4\times C_3$ is broken when the flavon fields $\varphi_T$, $\varphi_S$, $\xi$ transforming as $\bs{3}$, $\bs{3'}$, $\bs{1''}$, respectively, acquire the \vevs{} $\langle\varphi_T\rangle = (v_T, v_T, v_T), \enspace \langle\varphi_S\rangle = (v_S, 0, 0), \enspace \langle\xi\rangle = v_\xi$. The numbering of the subgroups corresponds to the output of the \gap{} script on p.~\pageref{verb:symbreak}.}
\label{tab:altferug-symbreak}
\end{table}

Note that since we will extend this analysis to 76 groups, we have to work with the generators that are supplied by \gap{}. It is not feasible to look for the optimal set of generators for each of the 76 groups that we will consider; in any case, the results are the same, and that the intermediate expressions may be more complicated is not relevant, since we have automated the calculation.

\medskip

We have written \python{} programs that interact with \gap{} to get the generators, the character table, the dimensions of the conjugacy classes and the explicit form of the matrices for all representations. From this, our code builds the Lagrangian that is invariant under all the symmetries, breaks the family symmetry, collects the terms that contribute to the charged lepton and neutrino mass matrices, and finally calculates the mixing matrix, the mixing angles and the phases. In \ref{sec:sym_lag_mix}, we will explain the details of our systematic scan.

\section{Systematic Construction of the Models}
\label{sec:sym_lag_mix}

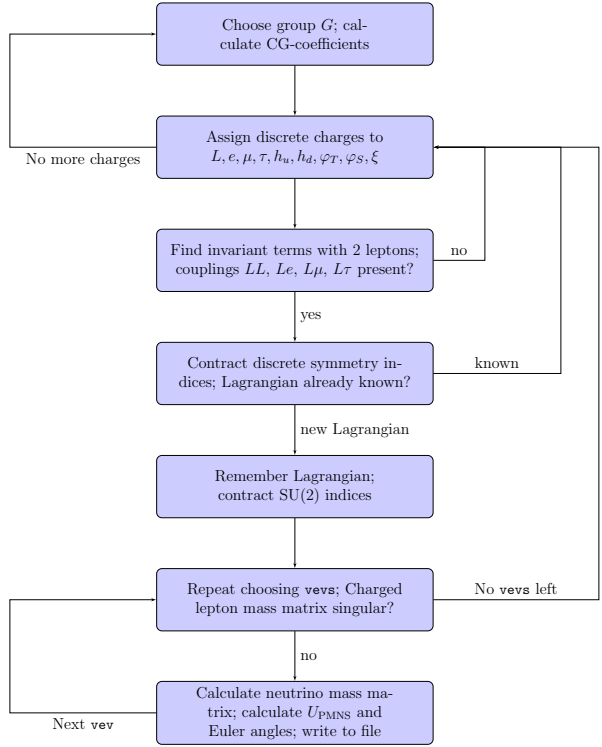
\begin{figure}[h!]
\centering
\scalebox{0.5}{
\begin{tikzpicture}[auto]
    \node [minimum width=7cm, block, text width=7cm] (B1) {Choose group $G$; calculate CG-coefficients};
    \node [minimum width=7cm, block, text width=7cm, node distance=3cm, below of=B1] (B2) {Assign discrete charges to $L,e,\mu,\tau,h_u,h_d,\varphi_T,\varphi_S,\xi$};
    \node [minimum width=7cm, block, text width=7cm, node distance=3cm, below of=B2] (B3) {Find invariant terms with 2 leptons; couplings $LL$, $Le$, $L\mu$, $L\tau$ present?};
    \node [minimum width=7cm, block, text width=7cm, node distance=3cm, below of=B3] (B5) {Contract discrete symmetry indices; Lagrangian already known?};
    \node [minimum width=7cm, block, text width=7cm, node distance=3cm, below of=B5] (B6) {Remember Lagrangian; contract \SU{2} indices};
    \node [minimum width=7cm, block, text width=7cm, node distance=3cm, below of=B6] (B7) {Repeat choosing \vevs{}; Charged lepton mass matrix singular?};
    \node [minimum width=7cm, block, text width=7cm, node distance=3cm, below of=B7] (B8) {Calculate neutrino mass matrix; calculate \UPMNS{} and Euler angles; write to file};
    \path [line] (B1) -- (B2);
    \path [line] (B2) -- (B3);
    \path [line] (B2) -- node {No more charges} +(-7.5,0) -- +(-7.5,3) -- (B1);
    \path [line] (B3) -- node {yes} (B5);
    \path [line] (B3) -- node {no} +(+5,0) -- +(+5,3) -- (B2);
    \path [line] (B5) -- node {known} +(+7,0) -- +(+7,6) -- (B2);
    \path [line] (B5) -- node {new Lagrangian} (B6);
    \path [line] (B6) -- (B7);
    \path [line] (B7) -- node {No \vevs{} left} +(+8,0) -- +(+8,12) -- (B2);
    \path [line] (B7) -- node {no}  (B8);
    \path [line] (B8) -- node {Next \vev{}} +(-7.5,0) -- +(-7.5,3) -- (B7);
\end{tikzpicture}}
\setcapindent{0em}
\caption{Systematic scan for models with family symmetry $G$ and up to three flavon fields.}
\label{flow:modelscan}
\end{figure}

In the following, we will consider the most general lepton sector with Standard Model particle content and up to three flavon fields. For clarity, we summarize our approach in form of a flow chart in \vref{flow:modelscan}  and elaborate on the details in \ref{sec:particlecontent}-\ref{sec:euler_angles_and_recognizing_tbm}.

\subsection{Particle Content}
\label{sec:particlecontent}

The particles and their Standard Model charges are listed in the first two columns of \vref{tab:particle_content} and will not be reproduced here. To avoid any misunderstandings, we emphasize that from now on $\xi$ is on the same footing as $\varphi_T$ and $\varphi_S$ and that its naming is simply a relic from earlier sections.

We restrict ourselves to such models where the lepton doublet $L$ transforms in a 3-dimensional representation and $e$, $\mu$, $\tau$ transform in 1-dimensional representations of the family symmetry. Plausible as this may sound, there is no physics reason for that, but rather, as we will explain below, without these assumptions the number of family charge assignments quickly grows too large to allow for a systematic scan. 

Regarding the Higgs sector, we will assume that there are exactly two fields. For one thing, we have supersymmetric models in mind that require an even number of Higgses. For another, more than two Higgs fields would spoil the unification of the gauge couplings. Thus, $h_u$, $h_d$ are assigned any 1-dimensional representation.

\subsection{Family Symmetry}

In the list of all groups of order $\leq100$ \cite{aw:2010:symmetries}, we find 90 groups which have a 3-dimensional representation (see \vref{tab:smallgroups_with_3dirrep}). We iterate over 76 out of these 90 groups that can be scanned in less than 60 days. Note that abelian groups only have 1-dimensional representations, and are thus not included in our scan. For a systematic scan of discrete \textit{abelian} symmetries, see ref.~\cite{Plentinger:2008up}. At this point in the algorithm, we calculate the relevant information on the group that we will need later on: 
\begin{inparaenum}[(i)]
\item The dimension of the group, the number and dimensions of its conjugacy classes, and its character table,
\item the irreducible representations and their tensor products,
\item the representation matrices for the irreducible representations,
\item the Clebsch-Gordan coefficients depending on the choice of the representation matrices.
\end{inparaenum}

\subsection{Charge Assignments}
\label{sec:charge_assignments}

We iterate over the inequivalent family charge assignments. As mentioned before, $L$ is assigned any 3-dimensional representation, and $e$, $\mu$, $\tau$, $h_u$, $h_d$ are assigned any 1-dimensional representations. We do not make any assumptions on the representations of the flavon fields $\varphi_T$, $\varphi_S$, $\xi$.

\medskip

In absence of a mechanism for generating mass hierarchies, we cannot distinguish between $e$, $\mu$ and $\tau$; they have the same quantum numbers and their naming is largely a matter of convention. Thus, to avoid iterating over configurations that give the same physics, we consider any permutation of the charge assignments to the $e$, $\mu$, $\tau$ to be equivalent. Once the mixing matrix has been derived, we can reorder its rows to recover the cases corresponding to the aforementioned permutations. In other words, we identify the electron, muon and tau \textit{a posteriori} and rename them where necessary.

The same holds for the flavon fields $\varphi_T$, $\varphi_S$, $\xi$, and since their interactions are not directly observable, it is not even necessary to rename them.

\medskip

The running time for the algorithm scales with the number of irreducible representations. Let us denote by $N_1$, $N_3$ and $N_a$ the number of 1-dimensional, 3-dimensional and all irreducible representations, respectively. Then the total number of inequivalent family charge assignments is
\begin{equation}
N_3 \enspace\times\enspace \mathcal{C}_r(N_1,3) \enspace\times\enspace N_1\cdot N_1 \enspace\times\enspace \mathcal{C}_r(N_a,3),
\end{equation}
where the first factor corresponds to $L$, the second one to $e$, $\mu$, $\tau$, the third one to $h_u$, $h_d$, and the last one to $\varphi_T$, $\varphi_S$, $\xi$. Consider e.g.~the second factor. The charge assignments to $e$, $\mu$, $\tau$ do not depend on their order, so if all three charges are distinct, the number of inequivalent choices is given by $N_1$-choose-3. In the more general case, two or more charges may be the same, and the number of inequivalent choices is given by (for $N=N_1$ and $k=3$):

\begin{equation}
\mathcal{C}_r(N,k) \equiv \mathcal{C}(N+k-1,k) = \frac{(N+k-1)!}{(N-1)!\,k!}
\end{equation}

The group $A_4\times C_3$ has $N_a=12$ irreducible representations with $N_3=3$ and $N_1=9$. Thus, the total number of inequivalent charge assignments is 14,594,580.

\medskip

Note that the case of less than three flavons is automatically included in the algorithm, since for a given flavon field we are also iterating over the \vev{} $v=0$ which effectively removes the corresponding field from the Lagrangian.

\subsection{Invariant Lagrangian}

In principle we could now construct the most general Lagrangian that is invariant under the gauge and family symmetries and contract the family indices. As a matter of fact, that is what we had initially done. However, because of the large number of inequivalent charge assignments, it is a better approach to try to determine as early in the algorithm as possible whether a given model is viable or not. 

For deriving the mass matrices of the charged and neutral leptons, we only need those terms in the Lagrangian that contain exactly 2 leptons, at most 2 Higgses and at most 1 flavon, since terms that have mass dimension greater than 6 or that are quadratic in the flavon fields are suppressed. We establish the invariance of a given term under the (gauge or family) symmetry by checking whether the tensor product of the particle representations contains a singlet. Note that this operation is very ``cheap'' for the computer as compared to doing the full contractions.

If there are no invariant terms at all or some of the couplings that we need for giving masses to the leptons are absent, we can skip the rest of the calculation and immediately continue with the next assignment of family charges. It is important to note that such improvements to the algorithm are crucial for keeping the running time within reasonable limits.

\subsection{Contracting the Indices}

To contract the family indices, we use the Clebsch-Gordan coefficients that we have already calculated in the first step of the algorithm. Their derivation for an arbitrary finite group (and choice of representation matrices) is not well-known, and the results available in the literature cover only specific cases. 

One algorithm for the general case that we are aware of was presented in ref.~\cite{Ludl:2009ft}. We have implemented an algorithm due to van den Broek and Cornwell \cite{vandenbroek:1978aa} that we believe to be more efficient and that is discussed \ref{app:clebsch-gordan-coefficients}.

After all gauge and family indices have been contracted, we rearrange the terms in the Lagrangian and the particles in each term to bring them into lexicographical order. We then compare the Lagrangian at hand with the list of Lagrangians from previous iterations. If the given Lagrangian is already known, we continue with the next iteration over family charges. Otherwise, we save it to the list and contract the \SU{2} indices. For $A_4\times C_3$, the number of different Lagrangians is 39,900.

\subsection{Substituting the Vacuum Expectation Values}
\label{sec:substituting_vevs}

In \ref{sec:breaking_the_family_symmetry} we show how to find all \vevs{} that break to a particular subgroup of the flavor symmetry. Unfortunately, we have examples which show that two different sets of \vevs{} that induce the same symmetry breaking pattern can lead to different values for the mixing angles, so our classification of the \vevs{} does not help us in classifying the models.

In lack of a better approach, we choose the entries of the \vevs{} to be 0 or 1. For $A_4\times C_3$, the number of different \vev{} configurations per family charge assignment may range from $2^3$ to $2^9$ ($\varphi_T$, $\varphi_S$, $\xi$ transform all in a 1-dimensional or 3-dimensional representation, respectively). Then we replace all fields by their vacuum expectation values. Analogously, we substitute $h_u^{(2)}=v_u$, $h_d^{(1)}=v_d$ and $h_u^{(1)}=h_d^{(2)}=0$, where the superscripts denote the \SU{2} indices.

\subsection{Mass and Mixing Matrices}

By construction, the Lagrangian contains only terms quadratic in the lepton fields whose coefficients give the charged and neutral lepton mass matrices $M_{\ell^+}$ and $M_{\nu}$, see \vref{eq:mass_matrices}. The singular value decomposition diagonalizes the mass matrices by unitary transformations whose product give the neutrino mixing matrix:
\begin{equation}
\hat{M}_{\ell^+} = D_L M_{\ell^+} D_R^\dagger, \qquad \hat{M}_{\nu} = U_L M_{\nu} U_R^\dagger, \qquad \UPMNS \equiv \DL \ULD
\end{equation}
If the charged or neutral lepton mass matrix is singular (i.e.~at least one of the masses is zero), we continue with the next iteration over the \vevs{}. Only when all of the \vevs{} are exhausted, we continue with the next iteration over the charge assignments.

\subsection{Euler Angles and Recognizing Tribimaximal Mixing}
\label{sec:euler_angles_and_recognizing_tbm}

Of the many different parametrizations \cite{Fritzsch:2001ty} for the neutrino mixing matrix that are mathematically equivalent and describe the same physics, we follow the standard notation as advocated by the \textit{Particle Data Group} \cite{Amsler:2008zzb}. To extract the mixing angles and phases, we use the explicit formulae presented in ref.~\cite{Plentinger:2006nb} that use a slightly different convention for the two Majorana phases.

\medskip

Note that $\UPMNS = \DL \ULD$ is guaranteed to be unitary by virtue of the singular value decomposition, but (before using the rephasing freedom) may not necessarily be in the standard form as given in ref.~\cite{Amsler:2008zzb}. Luckily, the formulae in ref.~\cite{Plentinger:2006nb} are applicable for any unitary matrix so that we can circumvent this technical complication. 

\medskip

As explained before, for a given charge assignment to $e$, $\mu$, $\tau$, we do not iterate over all its permutations, since we have the freedom of renaming the particles. As a consequence, when calculating the mixing angles, we must consider all permutations of the rows of $\UPMNS$. The naming of the neutrinos is then fixed by the corresponding charged leptons.

\subsection{Details of the Technical Implementation}

The programs were written in \code{Python 2.6.2} \cite{Python} and used \code{Numpy 1.4.1} \cite{oliphant:10} for the linear algebra operations. For the group theory calculations, we interfaced our programs with \code{GAP 4.4.12} \cite{GAP4}. For data analysis we used \code{PyROOT} \cite{Lavrijsen:pyroot2004} that provides a convenient interface to \code{Root 5.17} \cite{Brun:1997pa}. For generating the graphs, we mainly relied on \code{rootplot} \cite{Klukas:rootplot2010}, but in some cases we had to extend its functionality by overloading its classes and directly using \code{MatPlotLib 1.0.0} \cite{Hunter:2007}. The code was executed on the Linux cluster at the \emph{Centre de Calcul de l'Institut National de Physique Nucl\'{e}aire et Physique des Particules} in Lyon, France.

After the first initialization (creating the data on the group, opening files for reading and writing, etc.), it takes less than 1 second to calculate the full details for the model corresponding to the irrep assignments in \ref{sec:altarelli-feruglio-model} and 128 choices of \vevs{}. As expected, we rediscover tribimaximal mixing (for six \vev{} configurations).

\subsection{Running Times for the Different Groups}
\label{sec:runningtimes}

For $A_4\times C_3$, the execution time was 16 hours and 43 minutes on one computer with a 3 GHz Intel Xeon processor. From this, we get a useful measure to assess the running times for the other groups in \vref{tab:smallgroups_with_3dirrep}, since we know the number of inequivalent charge assignments for each group before we start running the programs. Note, however, that this can give us only an order of magnitude, because the computer center does not guarantee the same hardware on all of its machines.

The running times for the 90 groups of order $\leq100$ that have a 3-dimensional representation ranges from 29 seconds to ca.~177 years, whereas they do not necessarily increase with the number of elements in the group, but depend on the number of representations. For the present publication, we have decided to limit ourselves to those 76 groups that can be scanned in less than 60 days \textit{on one computer} and have indicated the 14 groups that exceed our time limit by red text color in \ref{tab:smallgroups_with_3dirrep}. For the actual calculations, we have distributed the calculations \textit{on more than one computer}.

It should be noted, though, that it is not impossible to tackle the groups that require longer running times. For one thing, one can add more theoretical priors to reduce the number of configurations that need to be scanned. For another, one can rewrite time-critical components in \cpp{} and integrate them into the \python{} programs. Yet another way to improve the code is to use parallel computing.

\section{Phenomenology}
\label{sec:analysis}

We will now discuss the results that we obtained from the systematic scan of family symmetries, charge assignments and vacuum configurations. We refer the reader also to \vref{tab:smallgroups_with_3dirrep} where some of the results of this section will be summarized.

It is important to stress that we are \textit{not} specifically searching for tribimaximal mixing, but constructing \textit{all models} for a given symmetry group (with the qualifications detailed in \ref{sec:charge_assignments} and \ref{sec:substituting_vevs}).

We will only list inequivalent models: We consider two models to be equivalent, if their Lagrangians after contracting the family indices, but before symmetry breaking are equal. In the plots, however, the data points correspond to vacua and some may correspond to the \textit{same} Lagrangian.

Since we will be discussing many different groups that may not all have a standard name, we will use their \gapid{}s, and e.g.~denote $A_4\times C_3$ by \gid{36}{11}. The correspondence between the \gapid{}s and the groups is given in \vref{tab:smallgroups_with_3dirrep}.

\subsection[Models for $A_4\times C_3$]{Models for $\bs{A_4\times C_3}$}

We will start with the results for the ``classic'' group $\gid{36}{11}=A_4\times C_3$. The 14,594,580 family charge assignments to $L,e,\mu,\tau,h_u,h_d,\varphi_T,\varphi_S,\xi$ give 39,900 inequivalent Lagrangians out of which 22,932 have non-singular charged lepton and neutrino mass matrices.

In this set,we find 4,233 models of tribimaximal mixing (18.5\%). For a given model, there may exist more than one vacuum configuration that leads to TBM (e.g.~6 \vevs{} for the model in \ref{sec:altarelli-feruglio-model}), and we have \textit{not} counted them separately.

\begin{figure}[h!]
\centering
\subfigure[\footnotesize Number of models that give $\theta_{ij}$ with no constraints on the other 2 angles. Each histogram has 15992118 entries.]{
\includegraphics[width=1.0\textwidth]{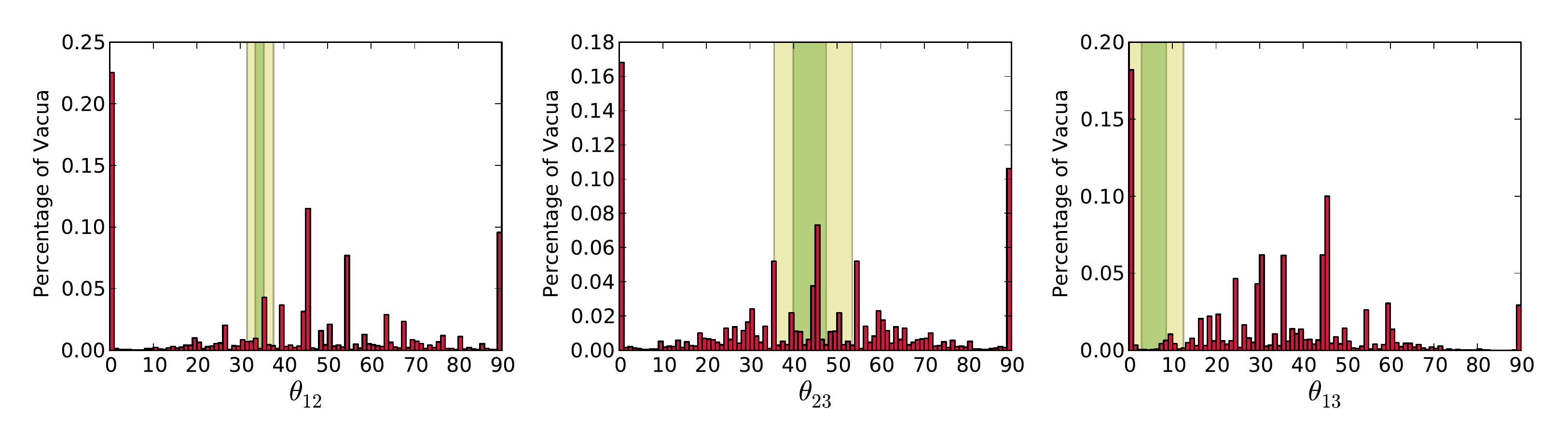}
\label{fig:angles-histo-1d-a}
}
\subfigure[Number of models that give $\theta_{ij}$ with the other 2 angles restricted to their $3\sigma$ interval. The histograms have 838289, 148886 and 225844 entries, respectively.]{
\includegraphics[width=1.0\textwidth]{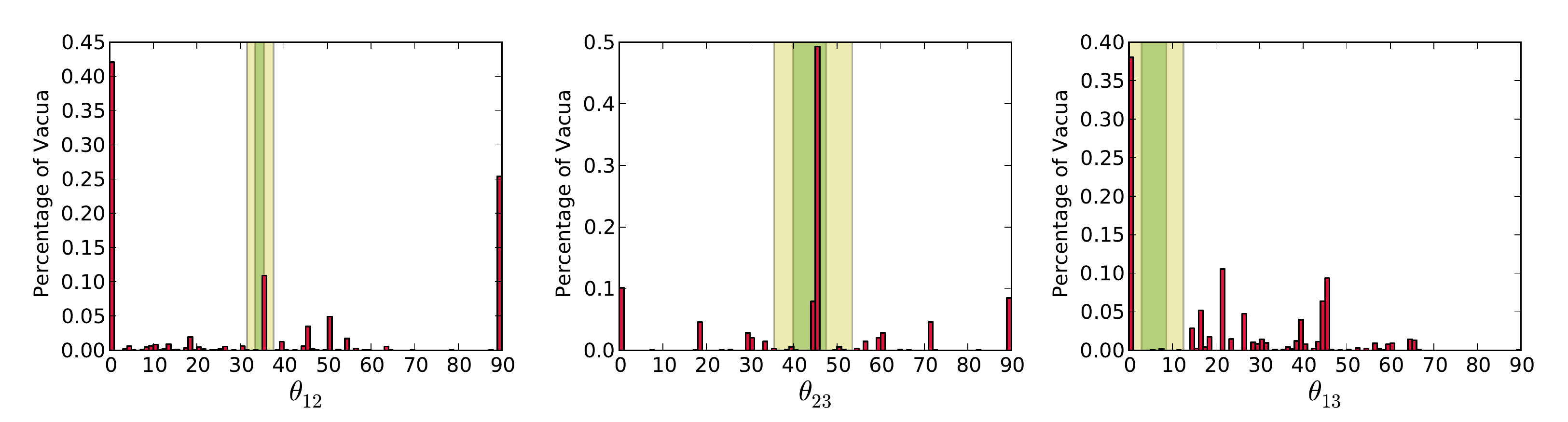}
\label{fig:angles-histo-1d-b}
}
\setcapindent{0em}
\caption{Number of models with family symmetry $\gid{36}{11}=A_4\times C_3$ that give the mixing angles denoted on the $x$-axis. The area of the histograms is normalized to 1 and the bin width is 1. The green and yellow bands correspond to the $1\sigma$ and $3\sigma$ ranges, respectively.}
\label{fig:angles-histo-1d}
\end{figure}

4,481 models (19.5\%) lie in the $3\sigma$ range of their measured values. We find no models that lie in the $1\sigma$ range, because $\theta_{13} = 0^\circ$ is excluded at $1\sigma$ (cf.~\vref{tab:experimental_data}). This fact is nicely illustrated in the third plot of \ref{fig:angles-histo-1d-b}: The $1\sigma$ range, represented by the green band, is empty.

\medskip

\ref{fig:angles-histo-1d} shows the distribution of the mixing angles $\theta_{12}$, $\theta_{23}$, and $\theta_{13}$, where we are now counting the \textit{vacua} and not the models. The reason for this is that for one and the same Lagrangian, the values of the mixing angles will in general depend on the choice of vacuum. The histograms in \ref{fig:angles-histo-1d-a} have 15,992,118 entries reflecting the fact that for each of the 39,900 inequivalent Lagrangians, we are looping over 8 to 512 vacua, depending on the dimensions of the irreps assigned to the flavon fields.

The area of each histogram has been normalized to 1 and the bin width is $1^\circ$, so the $y$-axis gives the \textit{percentage} of vacua that produce the angles on the $x$-axis. The green and yellow bands correspond to the $1\sigma$ and $3\sigma$ ranges, respectively (cf.~\vref{tab:experimental_data}). In \ref{fig:angles-histo-1d-a}, we simply count the number of times that a given angle is reproduced irrespective of the values that the two other angles may take. E.g.~from the first histogram we can read off that 7.6\% of the vacua give a value for $\theta_{12}$ that is consistent with experiment at $3\sigma$, where $\theta_{23}$ and $\theta_{13}$ can take any values.

\medskip

Now we investigate whether we can obtain some predictions by introducing priors. In \ref{fig:angles-histo-1d-b} we have restricted 2 of the angles to their $3\sigma$ intervals and plotted the third one. As a consequence, the numbers of entries in the histograms are not equal.

The most striking difference between \ref{fig:angles-histo-1d-a} and \ref{fig:angles-histo-1d-b} is that now the most likely value for $\theta_{23}$ is $45^\circ$ and lies in the $1\sigma$ interval. Furthermore, the number of vacua in the experimentally disfavored region has decreased significantly, and 58\% of the vacua are in the $3\sigma$ interval.

For $\theta_{13}$, values near $90^\circ$ are now excluded, and the $3\sigma$ interval is almost depopulated except for $\theta_{13}=0^\circ$, which at the same time corresponds to the maximum of the histogram. This can be interpreted as a prediction for $\theta_{13}$ to be $0^\circ$ (at leading order) based on current experimental data and the theoretical assumption of an $A_4$ family symmetry. 38\% of the vacua are in the $3\sigma$ interval.

For $\theta_{12}$, the most likely value is still $0^\circ$, but $35^\circ$ is now the third-most assumed angle. Clearly, the experimental data on $\theta_{23}$ and $\theta_{13}$ is pushing us in the right direction. 11\% of the vacua are in the $3\sigma$ interval.

It is an interesting observation that the effect of our ``cuts'' were such that the preference of the data for the experimentally allowed ranges became much more pronounced. This is an unexpected and non-trivial result and may further testify to the phenomenological viability of $A_4$.

\medskip

\begin{figure}[h!]
\centering
\subfigure[\footnotesize Number of models that give $\theta_{ij}$ and $\theta_{mn}$ with no constraint on the remaining angle. Each histogram has 15992118 entries.]{
\includegraphics[width=1\textwidth,viewport=2.3in 0in 21in 4.7in, clip]{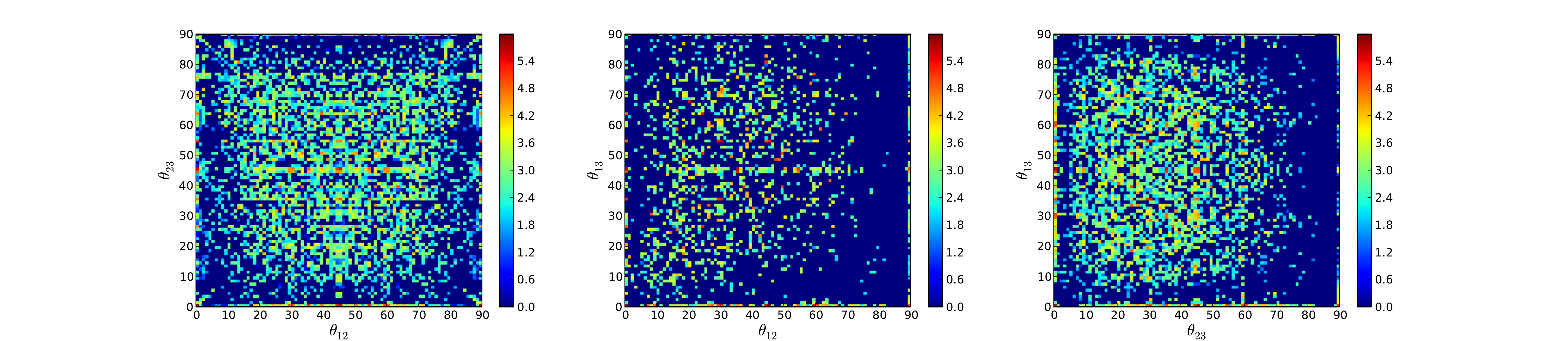}
\label{fig:angles-histo-2d-a}
}
\subfigure[\footnotesize Number of models that give $\theta_{ij}$ and $\theta_{mn}$ with the remaining angle restricted to its $3\sigma$ interval. The histograms have 2941000, 3675600 and 1057170 entries, respectively.]{
\includegraphics[width=1\textwidth,viewport=2.3in 0in 21in 4.7in, clip]{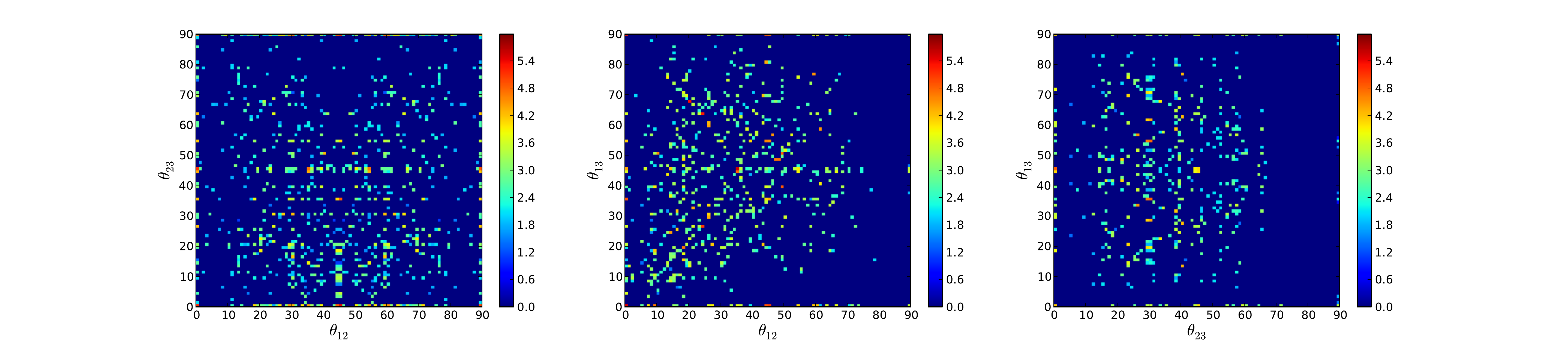}
\label{fig:angles-histo-2d-b}
}
\setcapindent{0em}
\caption{Logarithmic plot of the number of models with family symmetry $\gid{36}{11}=A_4\times C_3$ that give the mixing angles denoted on the axes. The bin width on both axes is 1. The base of the logarithm is 10, and the color map on the right side of each plot gives the exponents.}
\label{fig:angles-histo-2d}
\end{figure}

To learn more about the correlation of the angles and how priors may affect them, in \ref{fig:angles-histo-2d} we present the distribution of 2 out of 3 angles, respectively. The color bar on the right-hand side of each figure gives the correspondence between the colors in the plots and the logarithm to base 10 of the number of vacua with the angles $\theta_{ij}$ and $\theta_{mn}$ on the $x$- and $y$-axes, respectively. 

\medskip

In \ref{fig:angles-histo-2d-a}, each 2-dimensional histogram has 15,992,118 entries which correspond to the full set of vacua that we had also previously considered in \ref{fig:angles-histo-1d-a}. In analogy to \ref{fig:angles-histo-1d-a}, we have imposed no constraints on the third angle that is not plotted. From the first plot, we cannot read off much, except that there exist certain ``hot spots'' (e.g.~$\theta_{12}=0^\circ$ and $\theta_{23}=45^\circ$) that correspond to large numbers of vacua, and that the regions near the lower corners are by comparison less populated. In the second and third plots, we see that there are considerably fewer models for $\theta_{12}\gtrsim70^\circ$ and $\theta_{23}\gtrsim75^\circ$, respectively. In the case of the second plot, this holds even for much lower values of $\theta_{12}\gtrsim35^\circ$, given that $\theta_{13}$ is not larger than $\sim15^\circ$ or near $0^\circ$.

\medskip

In \ref{fig:angles-histo-2d-b}, we present the same correlations as in \ref{fig:angles-histo-2d-a}, but this time, we have required that the third angle be in its $3\sigma$ interval. As a consequence, the numbers of entries in the histograms are not equal. We have used the same normalization of the color bars in \ref{fig:angles-histo-2d-a} and \ref{fig:angles-histo-2d-b} to facilitate comparisons between them. 

Considering the first plot, we can see that most vacua lie in a band $\theta_{12}\simeq30^\circ-60^\circ$, whereas this effect becomes less pronounced for $\theta_{23}\simeq15^\circ-30^\circ$. For $\theta_{23}\simeq45^\circ$, values of $\theta_{12}=0^\circ,35^\circ,55^\circ,90^\circ$ are favored (red bins in plot).

The second plot clearly shows that $\theta_{12}\gtrsim70^\circ$ and to a lesser extent $\theta_{12}\lesssim10^\circ$ are disfavored. For $\theta_{12}\simeq10^\circ-70^\circ$, a band of $\theta_{13}=0^\circ-10^\circ$ that widens with increasing $\theta_{12}$ is sparsely populated, but note that for $\theta_{13}=0^\circ$ and $\theta_{12}=45^\circ$, there is one of the highest counts of vacua in the plot as indicated by the red bins. This is consistent with our previous observation from \ref{fig:angles-histo-1d-b} that $\theta_{13}=0^\circ$ is preferred, but that otherwise the region $\theta_{13}\lesssim10^\circ$ is disfavored.

In the third plot we again observe a band structure $\theta_{23}\simeq15^\circ-60^\circ$ where most of the vacua are concentrated, and find that $\theta_{13}\lesssim10^\circ$ and $\theta_{13}\gtrsim80^\circ$ are disfavored. The combination $\theta_{23}\simeq45^\circ$ and $\theta_{13}=0^\circ$, however, is not preferred. As indicated by the red bins, the most likely combination of angles lies elsewhere.

\begin{figure}[h!]
\centering
\subfigure[The 5,528 bins that are $\geq1$.]{
\includegraphics[width=0.47\textwidth]{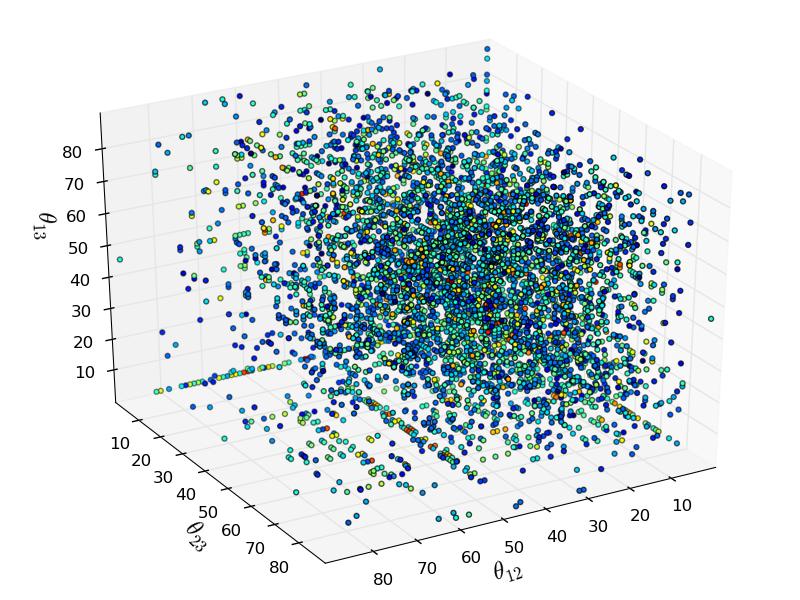}
\label{fig:angles-scatter-a}
}
\subfigure[The 1,287 bins that are $\geq1000$.]{
\includegraphics[width=0.47\textwidth]{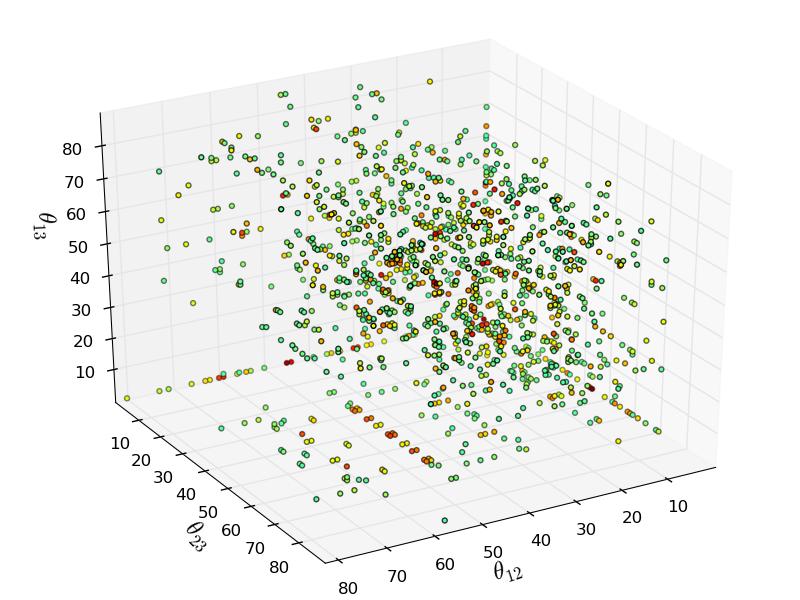}
\label{fig:angles-scatter-b}
}
\setcapindent{0em}
\caption{Scatter plot for $\theta_{12}$, $\theta_{23}$, $\theta_{13}$. Each point corresponds to a bin in the 3-dimensional histogram that has at least 1 entry. The bin width is 1. The color of the points (from blue to red) correspond to the logarithm of the number of models (from lower to higher).}
\label{fig:angles-scatter}
\end{figure}
 
Ideally, to give a graphical representation of the full information on the angles and their correlations, we would create a 3-dimensional histogram with $\theta_{12}$, $\theta_{23}$, and $\theta_{13}$ on the axes, and plot the number of vacua along a 4th dimension. Since this is not feasible, we present a plot in 3 dimensions, where the color of the data points indicates the number of vacua.

In \ref{fig:angles-scatter-a}, each point represents a bin in a 3-dimensional histogram: If there is at least one vacuum that produces the angles $(\theta_{12},\theta_{23},\theta_{13})$, we set a point at the respective coordinates. The bin width on each axis is 1, and in total there are $90\times90\times90$ bins, of which 5,528 are not empty. The color of the point denotes the logarithm to base 10 of the number of vacua that give the respective angles, where the colors from blue to red correspond to an increasing number of vacua. We have not displayed the color map for the plots, since we find it difficult to extract quantitative information from the 3-dimensional representation and rather use it as a means of uncovering correlations between the angles and the qualitative features of $A_4$ as a symmetry group.

In \ref{fig:angles-scatter-b}, we display only those 1,287 bins that have more than 1,000 entries. This removes much of the cluttering and gives a clearer picture regarding where the most likely vacua are concentrated. One immediate observation is that the parameter space for $(\theta_{12},\theta_{23},\theta_{13})$ is not uniformly populated: There are very few vacua for $\theta_{12}\simeq60^\circ-90^\circ$; low and high values of $\theta_{13}$ are disfavored (except $\theta_{13}=0^\circ$); and rotating \ref{fig:angles-scatter-b} to view it from different perspectives, we see that most of the vacua are concentrated in a volume $\theta_{12}\simeq10^\circ-60^\circ$, $\theta_{23}\simeq20^\circ-70^\circ$, $\theta_{13}\gtrsim15^\circ$.

Considering the $\theta_{12}-\theta_{23}$ plane of \ref{fig:angles-scatter-b} that corresponds to $\theta_{13}=0^\circ$, we find that $\theta_{12}=\theta_{23}=45^\circ$ are the most likely values (see red points), which is in agreement with the first plot in \ref{fig:angles-histo-2d-a}.

Regrettably, we fail to see any preference for tribimaximal mixing or the experimentally allowed values. Without putting in at least some priors, the best we can do is setting approximate upper and lower bounds on the angles.

\medskip

\begin{figure}[h!]
\centering
\subfigure[\footnotesize Number of models that give the phase denoted on the $x$-axis with no constraints on $\theta_{12}$, $\theta_{23}$, $\theta_{13}$. Each histogram has 15992118 entries.]{
\includegraphics[width=1.0\textwidth]{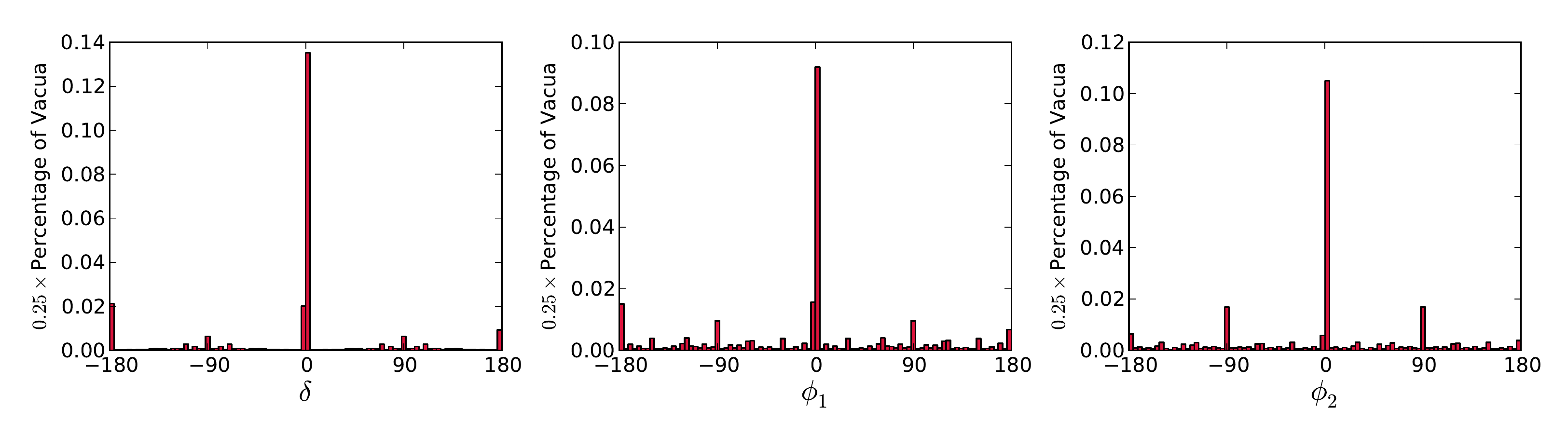}
\label{fig:phases-histo-1d-a}
}
\subfigure[\footnotesize Number of models that give the phase denoted on the $x$-axis with $\theta_{12}$, $\theta_{23}$, $\theta_{13}$ restricted to their $3\sigma$ intervals. Each histogram has 86014 entries.]{
\includegraphics[width=1.0\textwidth]{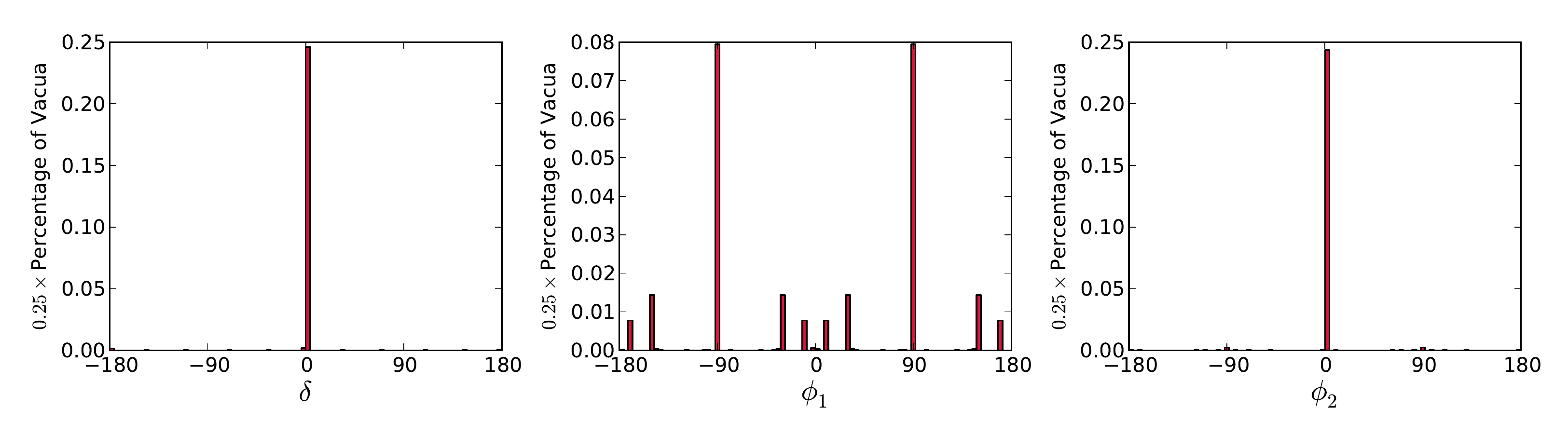}
\label{fig:phases-histo-1d-b}
}
\setcapindent{0em}
\caption{Number of models with family symmetry $\gid{36}{11}=A_4\times C_3$ that give the Dirac and Majorana phases denoted on the $x$-axis. The area of the histograms is normalized to 1 and the bin width is 4, so one must multiply the height of the bars by 4 to get the percentage of vacua.}
\label{fig:phases-histo-1d}
\end{figure}

If $\theta_{13}\neq0^\circ$, the CP-phase $\delta$ becomes relevant. In \ref{fig:phases-histo-1d}, we present the distribution of the vacua for $\delta$ and the two Majorana phases $\phi_1$ and $\phi_2$ (for our conventions, see ref.~\cite{Plentinger:2006nb}). Since the phases may take values from -$180$ to +$180$, we have chosen a bin width of 4. The area of the histograms is still normalized to 1, but now one has to multiply the height of the bars by a factor of 4 to get the percentage of vacua.

In \ref{fig:phases-histo-1d-a}, we simply count the number of vacua that realize the phase denoted on the $x$-axis, irrespective of the values that the other 2 phases and the 3 mixing angles may take. In \ref{fig:phases-histo-1d-b}, however, we use the experimental information that is available to us, namely, we restrict $\theta_{12}$, $\theta_{23}$ and $\theta_{13}$ to their respective $3\sigma$ intervals; experimental data on the phases is not available.

We get a clear prediction that $\delta=\phi_1=\phi_2=0$, which changes to $\delta=\phi_2=0$ and $\phi_1=\pm90$ if we take the experimental constraints into account. One reason for this may be that for the flavon fields, we have chosen only real vacuum expectation values. Complex numbers are introduced in the Lagrangian only through the Clebsch-Gordan coefficients.

In analogy to \ref{fig:angles-histo-2d} and \ref{fig:angles-scatter}, we have also analyzed the 2-dimensional histograms and the 3-dimensional scatter plots, but abstain from reproducing the graphs in the present publication. From the 2-dimensional histograms with the experimental constraints imposed, we learn that the phases prefer a discrete sets of values, and one may get hints at some correlations, e.g.~$\phi_2=\phi_1\pm180$. The 3-dimensional scatter plot is consistent with a random, uniform distribution of the phases, but we can distinguish some lines corresponding to a higher concentration of vacua.

\subsection[A Model with $\theta_{13}\simeq 5^\circ$]{A Model with $\bs{\theta_{13}\simeq 5}$}

Recently some analyses have reported on possible hints of a non-zero $\theta_{13}$ \cite{GonzalezGarcia:2010er,Mezzetto:2010zi,Fogli:2010zz}. A model with $0^\circ \lneqq \theta_{13} \leq 8.6^\circ$ that lies in the $1\sigma$ interval is easily constructed. For the family symmetry, we take $\gid{36}{11}=A_4\times C_3$ and assign the family charges $(L,e,\mu,\tau,h_u,h_d,\varphi_T,\varphi_S,\xi) \sim (\bs{3},\bs{1},\bsn{1}{4},\bs{1'''},\bs{1'},\bsn{1}{7},\bsn{1}{5},\bs{3''},\bs{3''})$. When the flavon fields acquire \vevs{} along the directions $\langle\varphi_{T}\rangle = (1)$, $\langle\varphi_{S}\rangle = (1,0,1)$, $\langle\xi\rangle = (1,1,1)$, we obtain the mixing angles $\theta_{12}=33.9^\circ$, $\theta_{23}=40.9^\circ$ and $\theta_{13}=5.1^\circ$ that all lie in the $1\sigma$ interval of the experimentally determined values. Incidentally, we have chosen the model in such a way as to produce a $\theta_{13}$ that is close to the present best-fit value of ref.~\cite{GonzalezGarcia:2010er} with the modified Gallium cross-section.

\subsection{Results for 76 Flavor Groups}

In this section, we present an overview of the models that we obtained from our systematic scan. Due to the sheer volume of the data, we will limit ourselves to making a few qualitative observations and relay the detailed analysis to an upcoming publication.

\begin{figure}[h!]
\addtolength{\abovecaptionskip}{-8mm}
\includegraphics[width=1.0\textwidth]{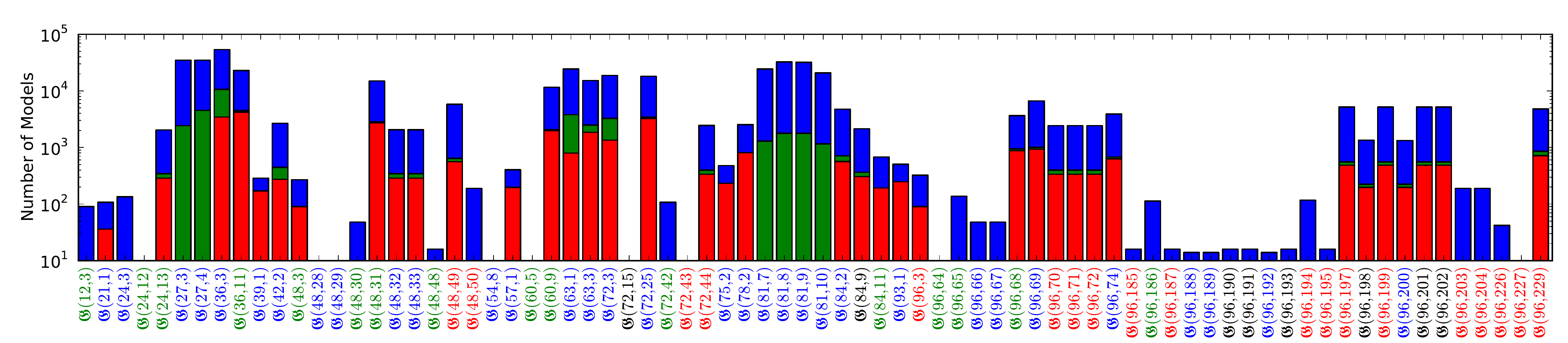}
\setcapindent{0em}
\caption{The number of models per symmetry group. On the $x$-axis, we label the flavor symmetry $\mathfrak{g}$ by its \gapid{}, cf.~\vref{tab:smallgroups_with_3dirrep}. The red and blue color of the labels on the $x$-axis indicates that $\mathfrak{g}\supset A_4$ and $\mathfrak{g}\subset \U{3}$, respectively, whereas the green color signifies that both conditions are satisfied simultaneously. Along the $y$-axis, the blue bars give the number of Lagrangians that lead to \textit{non-singular} mass matrices. The green bars indicates the number of models that lie withing the $3\sigma$ interval, and the red bars finally give the number of models for which at least one vacuum configuration gives tribimaximal mixing.}
\label{fig:numlagtbm}
\end{figure}

One main result of our analysis is that we have found thousands of new models that give exact tribimaximal mixing. \ref{fig:numlagtbm} shows the number of inequivalent models for each of the 76 groups that \begin{inparaenum}[(i)] \item are of order $\leq100$, \item have a 3-dimensional irreducible representation, and \item can be scanned in $\leq60$ days \end{inparaenum} (see \ref{sec:runningtimes} for more details). We have excluded those Lagrangians from our analysis that lead to a singular neutrino or charged lepton mass matrix. The red bars indicate the fraction of Lagrangians for which at least one choice of \vevs{} leads to tribimaximal mixing, and the blue bars give the number of models that lie in the $3\sigma$ interval of the measured angles (cf.~\vref{sec:experimental_results}). The correspondence between the \gapid{}s on the $x$-axis and the full name of the group is given in \vref{tab:smallgroups_with_3dirrep}, where we also list the exact numbers of  models that may be difficult to read off from the graph.

The conspicuous gaps in \ref{fig:numlagtbm} are a consequence of our criterion that the mass matrices be non-singular, i.e.~we do not consider such cases where any of the neutrinos (or charged leptons) is massless. 

Out of the 76 groups that we scanned, 9 (12\%) have only singular mass matrices. 44 groups (58\%) lie in the $3\sigma$ interval, and 38 (50\%) are even tribimaximal (for at least one vacuum configuration, respectively). Only for 23 groups (30\%) we could not find any vacuum configuration that satisfy the experimental limits. Note, though, that despite being very general, our scan is not fully comprehensive, since \begin{inparaenum}[(i)] \item we assume that the lepton doublet transforms in a triplet, and \item we do not scan over all possible \vevs{}. \end{inparaenum} Owing to this fact, there may exist even more viable models than we have identified.

To explore the connection between tribimaximal mixing and $A_4$, we have color-coded the group names on the $x$-axis of \ref{fig:numlagtbm}. The blue, red and green colors correspond to $\mathfrak{g}\subset \U{3}$, $\mathfrak{g}\supset A_4$, and $A_4 \subset \mathfrak{g}\subset \U{3}$, respectively. Of the 76 groups, 35 contain $A_4$ as a subgroup, but only for 16 out of these 35 groups we can find vacua that give models of tribimaximal mixing. It is conceivable, though, that one may find TBM models for the other groups, if one introduces more than 3 flavon fields. 

The chances of finding TBM does not scale with the total number of models, as e.g.~\gid{81}{7} and the next adjacent three groups show. Yet again, this conclusion may heavily depend on the number of flavon fields.

We find four groups, \gid{84}{9}, \gid{96}{198}, \gid{96}{201} and \gid{96}{202} that are neither subsets of \U{3} nor contain an $A_4$ subgroup and nevertheless can accommodate models of tribimaximal mixing.

\medskip

An interesting observation from \ref{fig:numlagtbm} is that for 9 groups the models with TBM are identical to those in the $3\sigma$ interval (the green bars are almost completely covered by the red ones). 5 of the 9 groups belong to the $T$-series of \SU{3} subgroups \cite{Fairbairn:1982jx}.

\begin{figure}[h!]
\addtolength{\abovecaptionskip}{-8mm}
\centering
\includegraphics[width=1.0\textwidth]{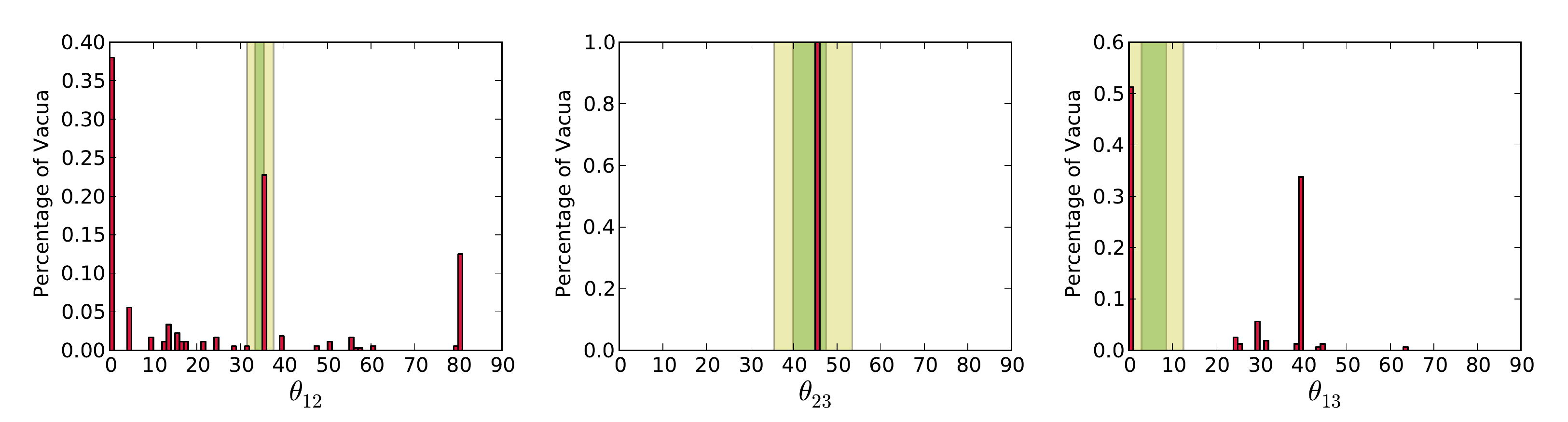}
\setcapindent{0em}
\caption{Number of models for $\gid{21}{1}=T_7$ that give $\theta_{ij}$ with the other 2 angles restricted to their $3\sigma$ interval. The histograms have 4472, 984 and 1920 entries, respectively. The area of the histograms is normalized to 1 and the bin width is 1. The green and yellow bands correspond to the $1\sigma$ and $3\sigma$ ranges, respectively.}
\label{fig:g21-1-angles-histo-1d}
\end{figure}

The smallest group for which we find TBM is $\gid{21}{1} = T_7$ which after $A_4$ is the smallest group that has a 3-dimensional irreducible representation. The family symmetry $T_7$ has been studied in refs.~\cite{Luhn:2007sy,Hagedorn:2008bc,Cao:2010mp}.

It is also worth mentioning that the second smallest group with models of TBM is $\gid{24}{13} = A_4\times C_2$ where the $C_3$ factor of the model in ref.~\cite{Altarelli:2005yx} has been replaced by a $C_2$.

\medskip

In \ref{fig:g21-1-angles-histo-1d} we show for $\gid{21}{1}=T_7$ the distribution of the mixing angles. If we use the experimental data on 2 of the angles and plot the multiplicity of the third one, we find that $\theta_{12}\simeq35^\circ$ is the second-most likely angle to be produced, and the only one within the $3\sigma$ interval. For $\theta_{23}$, we obtain a unique prediction $\theta_{23}=45^\circ$. As for $\theta_{13}$, the value $0^\circ$ is both the most likely angle as well as the only one attained within the $3\sigma$ interval.

\medskip

As compared to $\gid{36}{11}=A_4\times C_3$, there are other groups that have a larger fraction of TBM models. Consider e.g.~$\gid{39}{1}=C_{13} \rtimes_\varphi C_{3} = T_{13}$, where we find 288 inequivalent models, of which 171 (59\%) are TBM. Remarkably, any model for that group that has a vacuum for which the mixing angles are consistent with experiment at the $3\sigma$ level also allows for TBM to be realized.

\section{Conclusions}

In this publication we scanned 76 groups and constructed a total of 439,820 Lagrangians out of which 59,019 are consistent with experiment and 31,137 are tribimaximal. The large set of viable models allowed us to look at correlations between the mixing angles and make a prediction for $\theta_{13}$ that will be measured in upcoming experiments.

\smallskip

We have presented an explicit model with $\theta_{13}=5.1^\circ$ to show that the recent tentative hints of a non-zero $\theta_{13}$ can be accommodated. We found tribimaximal mixing in 38 flavor groups; most of these groups had not been considered for model building before. We hope that the calculational tools and methods that we have outlined will be useful for future model building efforts.

\smallskip

We would like to emphasize that we do not advocate a probabilistic approach to model building along the lines of the landscape idea in string theory. Rather, we are trying to maximize our chances of finding the correct model(s) by starting out with a large set that reproduces the mixing angles within the current experimental limits. In future, we plan to take this analysis several steps further and look at the generation of mass hierarchies, the vacuum alignment problem, and finally include the quark sector. Invariably, each step will drastically reduce the number of models, and the goal is to find at least one that passes all criteria.

\smallskip

On the other hand, for answering the question whether any discrete flavor group is inherently connected to tribimaximal mixing, a probabilistic approach may be useful: The easier tribimaximal mixing can be realized in a given group, the more pronounced is the connection. In this sense, $A_4$ fares well, but $T_{13}$ and maybe $T_7$ should be considered to be on the same footing, if not more promising.

\bigskip
\bigskip
\textbf{Acknowledgments} \\

\noindent
We acknowledge useful discussions with Christoph Luhn, Stuart Raby, and Sudhir Vempati. A.W.~would like to thank J\"org Meyer for drawing his attention to Python and pyRoot, and for always patiently answering his Root questions. We would like to thank the GAP Forum and in particular Vahid Dabbaghian, Alexander Hulpke, and Martin Sch\"onert for their support. A.W.~would like to thank the Korean Institute for Advanced Study (KIAS) and Eung-Jin Chun for their hospitality during the final stages of this work. K.P.~would like to thank the Indian Institute of Science and Sudhir Vempati and acknowledges support from the Department of Science and Technology of the Government of India under the grant  ``Complementarity from Direct and Indirect Searches of Supersymmetry''. We are greatly indebted to the \emph{Centre de Calcul de l'Institut National de Physique Nucl\'{e}aire et Physique des Particules} for using their resources. We thank Genevi\`eve B\'elanger, Gautam Bhattacharyya, J.F.~Cornwell, Amol Dighe, Sabine Kraml, Patrick Otto Ludl, Jong-Chul Park, Tzvetalina Stavreva for comments and suggestions.


\appendix

\labelformat{section}{Appendix #1} 
\addtocontents{toc}{\protect\setcounter{tocdepth}{1}}

\clearpage
\newpage
\section{List of Groups of Order at Most 100}
\label{app:90groups}

Only very few groups were given dedicated names by the mathematicians and physicists who studied them. Examples are the cyclic groups $C_n$, the symmetric groups $S_n$, the alternating groups $A_n$ and the dihedral groups $D_n$. The vast majority is described by their substructure and a ``prescription'' of how to put together these parts to form the full group. In \ref{app:elements_of_finite_group_theory} we sketch the concepts that are crucial for the description of the groups: the direct product, the semidirect product and the short exact sequence.

\bigskip

Since the list of 1,048 groups of order $\leq100$ is too long to include in the present publication, we have made it available for \href{http://lpsc.in2p3.fr/theorie/akin/documents/listof100smallgroups.pdf}{download} \cite{aw:2010:symmetries}. For details on the generation of this list, see \ref{sec:allgroups}.

\medskip

The list of the 90 groups of order $\leq100$ that have a 3-dimensional irreducible representation and that we have systematically scanned for viable models of lepton flavor is given in \ref{tab:smallgroups_with_3dirrep} below.

\subsection{Notation and Conventions}
\label{sec:notation-and-conventions-smallgroupnames}

Our notation follows the GAP Reference Manual p.~356 \cite{gap:2008:reference} with the following exceptions. We denote the direct product by ``$\times$'' and the semidirect by $N \rtimes_\varphi K$ where $N$ is normal. Beware that this convention is not unique and that the symbol ``$\rtimes_\varphi$'' may point the other way. In writing short exact sequences like $\bs{1} \rightarrow N  \rightarrow G \rightarrow Q \rightarrow \bs{1}$, we will omit the leading and trailing trivial groups in order to make our notation more compact.

\medskip

We denote the dihedral group of a regular $n$-gon by $D_n$, and {\it not} by $D_{2n}$, as some authors prefer to do. $C_n$ or $\mathbb{Z}_n$ is the cyclic group of order $n$. $S_n$ and $A_n$ are the symmetric and alternating groups, respectively. $Q_4$ and $Q_8$ are the quaternion and octonion groups, respectively. $\mathrm{SL}(n,p)$ is the special linear group over a finite field, i.e.~the set of all $n\times n$ matrices with determinant 1 and values from a field of order $p$.

To facilitate the comparison with the existing literature, we list in \ref{sec:smallgroups_alternative_names} a non-exhaustive list of alternative names for some of the groups considered in our analysis.

\medskip

Many of the groups that we consider do not have specific names, and we will refer to them by their \gapid{}s. \gid{m}{n} will denote the group that is generated in \gap{} by the command \code{SmallGroup($m$,$n$)}.

\subsection{The List of 90 Groups}

In \ref{tab:smallgroups_with_3dirrep} below, we list the 90 groups of order $\leq100$ that have a 3-dimensional irreducible representation. 

The first column gives the \gapid{} which is a label that uniquely identifies the group in \gap{}. The first number in the square brackets is the order of the group, and the second number simply enumerates different groups of the same order. The \gapid{}s of the 14 groups that require more than 60 days of computer time (see \ref{sec:runningtimes}) are marked in red. 

The second column gives the name of the group. If two or more groups by the same name are isomorphic, we list only one. For the conventions we used in naming the groups and for a non-exhaustive compilation of alternative names common in the physics and mathematics literature see \ref{sec:notation-and-conventions-smallgroupnames} and \ref{sec:smallgroups_alternative_names}, respectively. 

The third column indicates whether the group $G$ is a subgroup of \U{3}. If $G$ is in \SU{3}, an orange check mark is shown (\textcolor{BurntOrange}{\ding{52}}), otherwise a blue one (\textcolor{Blue}{\ding{52}}). 

The fourth and fifth columns indicate whether $G$ is in \U{2} or $\U{2}\times\U{1}$, respectively (blue check mark). If $G$ is in \SU{2} or $\SU{2}\times\U{1}$, respectively, the check mark is orange.

The sixth column indicates whether $G$ contains $A_4$ as a subgroup. The color of the check marks has no significance.

The seventh column gives the total number of inequivalent models, and the eight and ninth columns show the number of models that have vacua with mixing angles that lie in the $3\sigma$ interval or are tribimaximal, respectively.

\begin{center}
\footnotesize
\setcapindent{0em}
\rowcolors{1}{tableshade2}{white}
\renewcommand{\arraystretch}{1.4}
\newcolumntype{K}{>{\columncolor[gray]{.4}}c}
\newcolumntype{L}{>{\columncolor[gray]{.6}}c}
\newcolumntype{M}{>{\columncolor[gray]{.8}}c}
\newcolumntype{x}[1]{>{\centering\arraybackslash}p{#1}}
\begin{longtable}{|l|p{35ex}|c|c|c|c|r|r|r|}
\captionabove{The 90 groups of order $\leq100$ that have a 3-dimensional irreducible representation. For details, refer to the text preceding this table.}
\label{tab:smallgroups_with_3dirrep} 
\\

\hline \multicolumn{1}{|c|}{\gapid{}} & \multicolumn{1}{c|}{Group} & \multicolumn{1}{c|}{$U_3$} & \multicolumn{1}{c|}{$U_2$} & \multicolumn{1}{c|}{$U_{2\times1}$} & \multicolumn{1}{c|}{$A_4$} & \multicolumn{1}{|c|}{Models} & \multicolumn{1}{|c|}{$3\sigma$} & \multicolumn{1}{c|}{TBM} \\ 
\hline 
\endfirsthead

\multicolumn{9}{c}
{\tablename\ \thetable{} -- continued from previous page} \\
\hline \multicolumn{1}{|c|}{\gapid{}} & \multicolumn{1}{c|}{Group} & \multicolumn{1}{c|}{$U_3$} & \multicolumn{1}{c|}{$U_2$} & \multicolumn{1}{c|}{$U_{2\times1}$} & \multicolumn{1}{c|}{$A_4$} & \multicolumn{1}{|c|}{Models} & \multicolumn{1}{|c|}{$3\sigma$} & \multicolumn{1}{c|}{TBM} \\ 
\hline 
\endhead

\hline \multicolumn{9}{|r|}{{Continued on next page}} \\ \hline
\endfoot

\hline
\endlastfoot

$[ 12, 3 ] $ & $A_{4}$                                                      & \textcolor{BurntOrange}{\ding{52}} & \textcolor{SpringGreen}{\ding{55}} & \textcolor{SpringGreen}{\ding{55}} & \textcolor{Blue}{\ding{52}} & 90         & 0     & 0     \\
$[ 21, 1 ] $ & $C_{7} \rtimes_\varphi C_{3}$                                & \textcolor{BurntOrange}{\ding{52}} & \textcolor{SpringGreen}{\ding{55}} & \textcolor{SpringGreen}{\ding{55}} & \textcolor{SpringGreen}{\ding{55}} & 108        & 36    & 36    \\
$[ 24, 3 ] $ & $\text{SL}(2,3)$                                             & \textcolor{BurntOrange}{\ding{52}} & \textcolor{BurntOrange}{\ding{52}} & \textcolor{BurntOrange}{\ding{52}} & \textcolor{SpringGreen}{\ding{55}} & 135        & 0     & 0     \\
$[ 24, 12 ]$ & $S_{4}$                                                      & \textcolor{BurntOrange}{\ding{52}} & \textcolor{SpringGreen}{\ding{55}} & \textcolor{SpringGreen}{\ding{55}} & \textcolor{Blue}{\ding{52}} & 0          & 0     & 0     \\
$[ 24, 13 ]$ & $C_{2} \times A_{4}$                                         & \textcolor{Blue}{\ding{52}} & \textcolor{SpringGreen}{\ding{55}} & \textcolor{SpringGreen}{\ding{55}} & \textcolor{Blue}{\ding{52}} & 2034       & 344   & 288   \\
$[ 27, 3 ] $ & $(C_{3} \times C_{3}) \rtimes_\varphi C_{3}$                 & \textcolor{BurntOrange}{\ding{52}} & \textcolor{SpringGreen}{\ding{55}} & \textcolor{SpringGreen}{\ding{55}} & \textcolor{SpringGreen}{\ding{55}} & 34992      & 2430  & 0     \\
$[ 27, 4 ] $ & $C_{9} \rtimes_\varphi C_{3}$                                & \textcolor{Blue}{\ding{52}} & \textcolor{SpringGreen}{\ding{55}} & \textcolor{SpringGreen}{\ding{55}} & \textcolor{SpringGreen}{\ding{55}} & 34992      & 4536  & 0     \\
$[ 36, 3 ] $ & $(C_{2} \times C_{2}) \rtimes_\varphi C_{9}$                 & \textcolor{Blue}{\ding{52}} & \textcolor{SpringGreen}{\ding{55}} & \textcolor{SpringGreen}{\ding{55}} & \textcolor{SpringGreen}{\ding{55}} & 53535      & 10621 & 3459  \\
$[ 36, 11 ]$ & $C_{3} \times A_{4}$                                         & \textcolor{BurntOrange}{\ding{52}} & \textcolor{SpringGreen}{\ding{55}} & \textcolor{SpringGreen}{\ding{55}} & \textcolor{Blue}{\ding{52}} & 22932      & 4481  & 4233  \\
$[ 39, 1 ] $ & $C_{13} \rtimes_\varphi C_{3}$                               & \textcolor{BurntOrange}{\ding{52}} & \textcolor{SpringGreen}{\ding{55}} & \textcolor{SpringGreen}{\ding{55}} & \textcolor{SpringGreen}{\ding{55}} & 288        & 171   & 171   \\
$[ 42, 2 ] $ & $C_{2} \times (C_{7} \rtimes_\varphi C_{3})$                 & \textcolor{Blue}{\ding{52}} & \textcolor{SpringGreen}{\ding{55}} & \textcolor{SpringGreen}{\ding{55}} & \textcolor{SpringGreen}{\ding{55}} & 2682       & 445   & 273   \\
$[ 48, 3 ] $ & $(C_{4} \times C_{4}) \rtimes_\varphi C_{3}$                 & \textcolor{BurntOrange}{\ding{52}} & \textcolor{SpringGreen}{\ding{55}} & \textcolor{SpringGreen}{\ding{55}} & \textcolor{Blue}{\ding{52}} & 270        & 90    & 90    \\
$[ 48, 28 ]$ & $\text{SL}(2,3) \rightarrow G \rightarrow C_{2}$             & \textcolor{BurntOrange}{\ding{52}} & \textcolor{BurntOrange}{\ding{52}} & \textcolor{BurntOrange}{\ding{52}} & \textcolor{SpringGreen}{\ding{55}} & 0          & 0     & 0     \\
$[ 48, 29 ]$ & $\text{GL}(2,3)$                                             & \textcolor{BurntOrange}{\ding{52}} & \textcolor{Blue}{\ding{52}} & \textcolor{Blue}{\ding{52}} & \textcolor{SpringGreen}{\ding{55}} & 0          & 0     & 0     \\
$[ 48, 30 ]$ & $A_{4} \rtimes_\varphi C_{4}$                                & \textcolor{Blue}{\ding{52}} & \textcolor{SpringGreen}{\ding{55}} & \textcolor{SpringGreen}{\ding{55}} & \textcolor{Blue}{\ding{52}} & 48         & 0     & 0     \\
$[ 48, 31 ]$ & $C_{4} \times A_{4}$                                         & \textcolor{Blue}{\ding{52}} & \textcolor{SpringGreen}{\ding{55}} & \textcolor{SpringGreen}{\ding{55}} & \textcolor{Blue}{\ding{52}} & 14937      & 2864  & 2712  \\
$[ 48, 32 ]$ & $C_{2} \times \text{SL}(2,3)$                                & \textcolor{Blue}{\ding{52}} & \textcolor{SpringGreen}{\ding{55}} & \textcolor{BurntOrange}{\ding{52}} & \textcolor{SpringGreen}{\ding{55}} & 2052       & 344   & 288   \\
$[ 48, 33 ]$ & $\text{SL}(2,3) \rtimes_\varphi C_{2}$                       & \textcolor{BurntOrange}{\ding{52}} & \textcolor{Blue}{\ding{52}} & \textcolor{Blue}{\ding{52}} & \textcolor{SpringGreen}{\ding{55}} & 2052       & 344   & 288   \\
$[ 48, 48 ]$ & $C_{2} \times S_{4}$                                         & \textcolor{Blue}{\ding{52}} & \textcolor{SpringGreen}{\ding{55}} & \textcolor{SpringGreen}{\ding{55}} & \textcolor{Blue}{\ding{52}} & 16         & 0     & 0     \\
$[ 48, 49 ]$ & $C_{2} \times C_{2} \times A_{4}$                            & \textcolor{SpringGreen}{\ding{55}} & \textcolor{SpringGreen}{\ding{55}} & \textcolor{SpringGreen}{\ding{55}} & \textcolor{Blue}{\ding{52}} & 5805       & 640   & 561   \\
$[ 48, 50 ]$ & $(C_{2} \times C_{2} \times C_{2} \times C_{2}) \rtimes_\varphi C_{3}$ & \textcolor{SpringGreen}{\ding{55}} & \textcolor{SpringGreen}{\ding{55}} & \textcolor{SpringGreen}{\ding{55}} & \textcolor{Blue}{\ding{52}} & 189        & 0     & 0     \\
$[ 54, 8 ] $ & $((C_{3} \times C_{3}) \rtimes_\varphi C_{3}) \rtimes_\varphi C_{2}$ & \textcolor{BurntOrange}{\ding{52}} & \textcolor{SpringGreen}{\ding{55}} & \textcolor{SpringGreen}{\ding{55}} & \textcolor{SpringGreen}{\ding{55}} & 0          & 0     & 0     \\
$\Red{[ 54, 10 ]}$ & $C_{2} \times ((C_{3} \times C_{3}) \rtimes_\varphi C_{3})$  & \textcolor{Blue}{\ding{52}} & \textcolor{SpringGreen}{\ding{55}} & \textcolor{SpringGreen}{\ding{55}} & \textcolor{SpringGreen}{\ding{55}} & 0          & 0     & 0     \\
$\Red{[ 54, 11 ]}$ & $C_{2} \times (C_{9} \rtimes_\varphi C_{3})$                 & \textcolor{Blue}{\ding{52}} & \textcolor{SpringGreen}{\ding{55}} & \textcolor{SpringGreen}{\ding{55}} & \textcolor{SpringGreen}{\ding{55}} & 0          & 0     & 0     \\
$[ 57, 1 ] $ & $C_{19} \rtimes_\varphi C_{3}$                               & \textcolor{BurntOrange}{\ding{52}} & \textcolor{SpringGreen}{\ding{55}} & \textcolor{SpringGreen}{\ding{55}} & \textcolor{SpringGreen}{\ding{55}} & 405        & 198   & 198   \\
$[ 60, 5 ] $ & $A_{5}$                                                      & \textcolor{BurntOrange}{\ding{52}} & \textcolor{SpringGreen}{\ding{55}} & \textcolor{SpringGreen}{\ding{55}} & \textcolor{Blue}{\ding{52}} & 0          & 0     & 0     \\
$[ 60, 9 ] $ & $C_{5} \times A_{4}$                                         & \textcolor{Blue}{\ding{52}} & \textcolor{SpringGreen}{\ding{55}} & \textcolor{SpringGreen}{\ding{55}} & \textcolor{Blue}{\ding{52}} & 11575      & 2063  & 1983  \\
$[ 63, 1 ] $ & $C_{7} \rtimes_\varphi C_{9}$                                & \textcolor{Blue}{\ding{52}} & \textcolor{SpringGreen}{\ding{55}} & \textcolor{SpringGreen}{\ding{55}} & \textcolor{SpringGreen}{\ding{55}} & 24345      & 3792  & 795   \\
$[ 63, 3 ] $ & $C_{3} \times (C_{7} \rtimes_\varphi C_{3})$                 & \textcolor{BurntOrange}{\ding{52}} & \textcolor{SpringGreen}{\ding{55}} & \textcolor{SpringGreen}{\ding{55}} & \textcolor{SpringGreen}{\ding{55}} & 15246      & 2483  & 1863  \\
$[ 72, 3 ] $ & $Q_{8} \rtimes_\varphi C_{9}$                                & \textcolor{BurntOrange}{\ding{52}} & \textcolor{Blue}{\ding{52}} & \textcolor{BurntOrange}{\ding{52}} & \textcolor{SpringGreen}{\ding{55}} & 18714      & 3272  & 1344  \\
$[ 72, 15 ]$ & $((C_{2} \times C_{2}) \rtimes_\varphi C_{9}) \rtimes_\varphi C_{2}$ & \textcolor{SpringGreen}{\ding{55}} & \textcolor{SpringGreen}{\ding{55}} & \textcolor{SpringGreen}{\ding{55}} & \textcolor{SpringGreen}{\ding{55}} & 0          & 0     & 0     \\
$\Red{[ 72, 16 ]}$ & $C_{2} \times ((C_{2} \times C_{2}) \rtimes_\varphi C_{9})$  & \textcolor{Blue}{\ding{52}} & \textcolor{SpringGreen}{\ding{55}} & \textcolor{SpringGreen}{\ding{55}} & \textcolor{SpringGreen}{\ding{55}} & 0          & 0     & 0     \\
$[ 72, 25 ]$ & $C_{3} \times \text{SL}(2,3)$                                & \textcolor{BurntOrange}{\ding{52}} & \textcolor{Blue}{\ding{52}} & \textcolor{BurntOrange}{\ding{52}} & \textcolor{SpringGreen}{\ding{55}} & 18105      & 3441  & 3261  \\
$[ 72, 42 ]$ & $C_{3} \times S_{4}$                                         & \textcolor{BurntOrange}{\ding{52}} & \textcolor{SpringGreen}{\ding{55}} & \textcolor{SpringGreen}{\ding{55}} & \textcolor{Blue}{\ding{52}} & 108        & 0     & 0     \\
$[ 72, 43 ]$ & $(C_{3} \times A_{4}) \rtimes_\varphi C_{2}$                 & \textcolor{SpringGreen}{\ding{55}} & \textcolor{SpringGreen}{\ding{55}} & \textcolor{SpringGreen}{\ding{55}} & \textcolor{Blue}{\ding{52}} & 0          & 0     & 0     \\
$[ 72, 44 ]$ & $A_{4} \times S_{3}$                                         & \textcolor{SpringGreen}{\ding{55}} & \textcolor{SpringGreen}{\ding{55}} & \textcolor{SpringGreen}{\ding{55}} & \textcolor{Blue}{\ding{52}} & 2451       & 399   & 336   \\
$\Red{[ 72, 47 ]}$ & $C_{6} \times A_{4}$                                         & \textcolor{Blue}{\ding{52}} & \textcolor{SpringGreen}{\ding{55}} & \textcolor{SpringGreen}{\ding{55}} & \textcolor{Blue}{\ding{52}} & 0          & 0     & 0     \\
$[ 75, 2 ] $ & $(C_{5} \times C_{5}) \rtimes_\varphi C_{3}$                 & \textcolor{BurntOrange}{\ding{52}} & \textcolor{SpringGreen}{\ding{55}} & \textcolor{SpringGreen}{\ding{55}} & \textcolor{SpringGreen}{\ding{55}} & 477        & 234   & 234   \\
$[ 78, 2 ] $ & $C_{2} \times (C_{13} \rtimes_\varphi C_{3})$                & \textcolor{Blue}{\ding{52}} & \textcolor{SpringGreen}{\ding{55}} & \textcolor{SpringGreen}{\ding{55}} & \textcolor{SpringGreen}{\ding{55}} & 2541       & 810   & 810   \\
$\Red{[ 81, 3 ]}$ & $(C_{9} \times C_{3}) \rtimes_\varphi C_{3}$                 & \textcolor{SpringGreen}{\ding{55}} & \textcolor{SpringGreen}{\ding{55}} & \textcolor{SpringGreen}{\ding{55}} & \textcolor{SpringGreen}{\ding{55}} & 0          & 0     & 0     \\
$\Red{[ 81, 4 ]}$ & $C_{9} \rtimes_\varphi C_{9}$                                & \textcolor{SpringGreen}{\ding{55}} & \textcolor{SpringGreen}{\ding{55}} & \textcolor{SpringGreen}{\ding{55}} & \textcolor{SpringGreen}{\ding{55}} & 0          & 0     & 0     \\
$\Red{[ 81, 6 ]}$ & $C_{27} \rtimes_\varphi C_{3}$                               & \textcolor{Blue}{\ding{52}} & \textcolor{SpringGreen}{\ding{55}} & \textcolor{SpringGreen}{\ding{55}} & \textcolor{SpringGreen}{\ding{55}} & 0          & 0     & 0     \\
$[ 81, 7 ] $ & $(C_{3} \times C_{3} \times C_{3}) \rtimes_\varphi C_{3}$    & \textcolor{Blue}{\ding{52}} & \textcolor{SpringGreen}{\ding{55}} & \textcolor{SpringGreen}{\ding{55}} & \textcolor{SpringGreen}{\ding{55}} & 24329      & 1296  & 0     \\
$[ 81, 8 ] $ & $(C_{9} \times C_{3}) \rtimes_\varphi C_{3}$                 & \textcolor{Blue}{\ding{52}} & \textcolor{SpringGreen}{\ding{55}} & \textcolor{SpringGreen}{\ding{55}} & \textcolor{SpringGreen}{\ding{55}} & 32416      & 1782  & 0     \\
$[ 81, 9 ] $ & $(C_{9} \times C_{3}) \rtimes_\varphi C_{3}$                 & \textcolor{BurntOrange}{\ding{52}} & \textcolor{SpringGreen}{\ding{55}} & \textcolor{SpringGreen}{\ding{55}} & \textcolor{SpringGreen}{\ding{55}} & 32076      & 1782  & 0     \\
$[ 81, 10 ]$ & $(C_{3} \times C_{3}) \rightarrow G \rightarrow (C_{3} \times C_{3})$ & \textcolor{Blue}{\ding{52}} & \textcolor{SpringGreen}{\ding{55}} & \textcolor{SpringGreen}{\ding{55}} & \textcolor{SpringGreen}{\ding{55}} & 20736      & 1161  & 0     \\
$\Red{[ 81, 12 ]}$ & $C_{3} \times ((C_{3} \times C_{3}) \rtimes_\varphi C_{3})$  & \textcolor{SpringGreen}{\ding{55}} & \textcolor{SpringGreen}{\ding{55}} & \textcolor{SpringGreen}{\ding{55}} & \textcolor{SpringGreen}{\ding{55}} & 0          & 0     & 0     \\
$\Red{[ 81, 13 ]}$ & $C_{3} \times (C_{9} \rtimes_\varphi C_{3})$                 & \textcolor{SpringGreen}{\ding{55}} & \textcolor{SpringGreen}{\ding{55}} & \textcolor{SpringGreen}{\ding{55}} & \textcolor{SpringGreen}{\ding{55}} & 0          & 0     & 0     \\
$\Red{[ 81, 14 ]}$ & $(C_{9} \times C_{3}) \rtimes_\varphi C_{3}$                 & \textcolor{Blue}{\ding{52}} & \textcolor{SpringGreen}{\ding{55}} & \textcolor{SpringGreen}{\ding{55}} & \textcolor{SpringGreen}{\ding{55}} & 0          & 0     & 0     \\
$[ 84, 2 ] $ & $C_{4} \times (C_{7} \rtimes_\varphi C_{3})$                 & \textcolor{Blue}{\ding{52}} & \textcolor{SpringGreen}{\ding{55}} & \textcolor{SpringGreen}{\ding{55}} & \textcolor{SpringGreen}{\ding{55}} & 4752       & 714   & 567   \\
$[ 84, 9 ] $ & $C_{2} \times C_{2} \times (C_{7} \rtimes_\varphi C_{3})$    & \textcolor{SpringGreen}{\ding{55}} & \textcolor{SpringGreen}{\ding{55}} & \textcolor{SpringGreen}{\ding{55}} & \textcolor{SpringGreen}{\ding{55}} & 2136       & 366   & 306   \\
$\Red{[ 84, 10 ]}$ & $C_{7} \times A_{4}$                                         & \textcolor{Blue}{\ding{52}} & \textcolor{SpringGreen}{\ding{55}} & \textcolor{SpringGreen}{\ding{55}} & \textcolor{Blue}{\ding{52}} & 0          & 0     & 0     \\
$[ 84, 11 ]$ & $(C_{14} \times C_{2}) \rtimes_\varphi C_{3}$                & \textcolor{BurntOrange}{\ding{52}} & \textcolor{SpringGreen}{\ding{55}} & \textcolor{SpringGreen}{\ding{55}} & \textcolor{Blue}{\ding{52}} & 678        & 192   & 192   \\
$[ 93, 1 ] $ & $C_{31} \rtimes_\varphi C_{3}$                               & \textcolor{BurntOrange}{\ding{52}} & \textcolor{SpringGreen}{\ding{55}} & \textcolor{SpringGreen}{\ding{55}} & \textcolor{SpringGreen}{\ding{55}} & 507        & 249   & 249   \\
$[ 96, 3 ] $ & $((C_{4} \times C_{2}) \rtimes_\varphi C_{4}) \rtimes_\varphi C_{3}$ & \textcolor{SpringGreen}{\ding{55}} & \textcolor{SpringGreen}{\ding{55}} & \textcolor{SpringGreen}{\ding{55}} & \textcolor{Blue}{\ding{52}} & 324        & 90    & 90    \\
$[ 96, 64 ]$ & $((C_{4} \times C_{4}) \rtimes_\varphi C_{3}) \rtimes_\varphi C_{2}$ & \textcolor{BurntOrange}{\ding{52}} & \textcolor{SpringGreen}{\ding{55}} & \textcolor{SpringGreen}{\ding{55}} & \textcolor{Blue}{\ding{52}} & 0          & 0     & 0     \\
$[ 96, 65 ]$ & $A_{4} \rtimes_\varphi C_{8}$                                & \textcolor{Blue}{\ding{52}} & \textcolor{SpringGreen}{\ding{55}} & \textcolor{SpringGreen}{\ding{55}} & \textcolor{Blue}{\ding{52}} & 138        & 0     & 0     \\
$[ 96, 66 ]$ & $\text{SL}(2,3) \rtimes_\varphi C_{4}$                       & \textcolor{Blue}{\ding{52}} & \textcolor{SpringGreen}{\ding{55}} & \textcolor{BurntOrange}{\ding{52}} & \textcolor{SpringGreen}{\ding{55}} & 48         & 0     & 0     \\
$[ 96, 67 ]$ & $\text{SL}(2,3) \rtimes_\varphi C_{4}$                       & \textcolor{BurntOrange}{\ding{52}} & \textcolor{Blue}{\ding{52}} & \textcolor{Blue}{\ding{52}} & \textcolor{SpringGreen}{\ding{55}} & 48         & 0     & 0     \\
$[ 96, 68 ]$ & $C_{2} \times ((C_{4} \times C_{4}) \rtimes_\varphi C_{3})$  & \textcolor{Blue}{\ding{52}} & \textcolor{SpringGreen}{\ding{55}} & \textcolor{SpringGreen}{\ding{55}} & \textcolor{Blue}{\ding{52}} & 3648       & 939   & 876   \\
$[ 96, 69 ]$ & $C_{4} \times \text{SL}(2,3)$                                & \textcolor{Blue}{\ding{52}} & \textcolor{SpringGreen}{\ding{55}} & \textcolor{BurntOrange}{\ding{52}} & \textcolor{SpringGreen}{\ding{55}} & 6637       & 1002  & 936   \\
$[ 96, 70 ]$ & $((C_{2} \times C_{2} \times C_{2} \times C_{2}) \rtimes_\varphi C_{3}) \rtimes_\varphi C_{2}$ & \textcolor{SpringGreen}{\ding{55}} & \textcolor{SpringGreen}{\ding{55}} & \textcolor{SpringGreen}{\ding{55}} & \textcolor{Blue}{\ding{52}} & 2433       & 399   & 336   \\
$[ 96, 71 ]$ & $((C_{4} \times C_{4}) \rtimes_\varphi C_{3}) \rtimes_\varphi C_{2}$ & \textcolor{SpringGreen}{\ding{55}} & \textcolor{SpringGreen}{\ding{55}} & \textcolor{SpringGreen}{\ding{55}} & \textcolor{Blue}{\ding{52}} & 2433       & 399   & 336   \\
$[ 96, 72 ]$ & $((C_{4} \times C_{4}) \rtimes_\varphi C_{3}) \rtimes_\varphi C_{2}$ & \textcolor{SpringGreen}{\ding{55}} & \textcolor{SpringGreen}{\ding{55}} & \textcolor{SpringGreen}{\ding{55}} & \textcolor{Blue}{\ding{52}} & 2433       & 399   & 336   \\
$\Red{[ 96, 73 ]}$ & $C_{8} \times A_{4}$                                         & \textcolor{Blue}{\ding{52}} & \textcolor{SpringGreen}{\ding{55}} & \textcolor{SpringGreen}{\ding{55}} & \textcolor{Blue}{\ding{52}} & 0          & 0     & 0     \\
$[ 96, 74 ]$ & $((C_{8} \times C_{2}) \rtimes_\varphi C_{2}) \rtimes_\varphi C_{3}$ & \textcolor{BurntOrange}{\ding{52}} & \textcolor{Blue}{\ding{52}} & \textcolor{Blue}{\ding{52}} & \textcolor{SpringGreen}{\ding{55}} & 3884       & 678   & 630   \\
$[ 96, 185 ]$ & $A_{4} \rtimes_\varphi Q_{8}$                                & \textcolor{SpringGreen}{\ding{55}} & \textcolor{SpringGreen}{\ding{55}} & \textcolor{SpringGreen}{\ding{55}} & \textcolor{Blue}{\ding{52}} & 16         & 0     & 0     \\
$[ 96, 186 ]$ & $C_{4} \times S_{4}$                                         & \textcolor{Blue}{\ding{52}} & \textcolor{SpringGreen}{\ding{55}} & \textcolor{SpringGreen}{\ding{55}} & \textcolor{Blue}{\ding{52}} & 113        & 0     & 0     \\
$[ 96, 187 ]$ & $(C_{2} \times S_{4}) \rtimes_\varphi C_{2}$                 & \textcolor{SpringGreen}{\ding{55}} & \textcolor{SpringGreen}{\ding{55}} & \textcolor{SpringGreen}{\ding{55}} & \textcolor{Blue}{\ding{52}} & 16         & 0     & 0     \\
$[ 96, 188 ]$ & $\text{SL}(2,3) \rightarrow G \rightarrow C_{2})$            & \textcolor{Blue}{\ding{52}} & \textcolor{SpringGreen}{\ding{55}} & \textcolor{BurntOrange}{\ding{52}} & \textcolor{SpringGreen}{\ding{55}} & 14         & 0     & 0     \\
$[ 96, 189 ]$ & $C_{2} \times \text{GL}(2,3)$                                & \textcolor{Blue}{\ding{52}} & \textcolor{SpringGreen}{\ding{55}} & \textcolor{Blue}{\ding{52}} & \textcolor{SpringGreen}{\ding{55}} & 14         & 0     & 0     \\
$[ 96, 190 ]$ & $(C_{2} \times \text{SL}(2,3)) \rtimes_\varphi C_{2}$        & \textcolor{SpringGreen}{\ding{55}} & \textcolor{SpringGreen}{\ding{55}} & \textcolor{SpringGreen}{\ding{55}} & \textcolor{SpringGreen}{\ding{55}} & 16         & 0     & 0     \\
$[ 96, 191 ]$ & $\text{SL}(2,3) \rightarrow G \rightarrow C_{2}) \rtimes_\varphi C_{2}$ & \textcolor{SpringGreen}{\ding{55}} & \textcolor{SpringGreen}{\ding{55}} & \textcolor{SpringGreen}{\ding{55}} & \textcolor{SpringGreen}{\ding{55}} & 16         & 0     & 0     \\
$[ 96, 192 ]$ & $\text{SL}(2,3) \rightarrow G \rightarrow C_{2}) \rtimes_\varphi C_{2}$ & \textcolor{BurntOrange}{\ding{52}} & \textcolor{Blue}{\ding{52}} & \textcolor{Blue}{\ding{52}} & \textcolor{SpringGreen}{\ding{55}} & 14         & 0     & 0     \\
$[ 96, 193 ]$ & $(\text{SL}(2,3) \rtimes_\varphi C_{2}) \rtimes_\varphi C_{2}$ & \textcolor{SpringGreen}{\ding{55}} & \textcolor{SpringGreen}{\ding{55}} & \textcolor{SpringGreen}{\ding{55}} & \textcolor{SpringGreen}{\ding{55}} & 16         & 0     & 0     \\
$[ 96, 194 ]$ & $C_{2} \times (A_{4} \rtimes_\varphi C_{4})$                 & \textcolor{SpringGreen}{\ding{55}} & \textcolor{SpringGreen}{\ding{55}} & \textcolor{SpringGreen}{\ding{55}} & \textcolor{Blue}{\ding{52}} & 118        & 0     & 0     \\
$[ 96, 195 ]$ & $(C_{2} \times C_{2} \times A_{4}) \rtimes_\varphi C_{2}$    & \textcolor{SpringGreen}{\ding{55}} & \textcolor{SpringGreen}{\ding{55}} & \textcolor{SpringGreen}{\ding{55}} & \textcolor{Blue}{\ding{52}} & 16         & 0     & 0     \\
$\Red{[ 96, 196 ]}$ & $C_{2} \times C_{4} \times A_{4}$                            & \textcolor{SpringGreen}{\ding{55}} & \textcolor{SpringGreen}{\ding{55}} & \textcolor{SpringGreen}{\ding{55}} & \textcolor{Blue}{\ding{52}} & 0          & 0     & 0     \\
$[ 96, 197 ]$ & $D_{4} \times A_{4}$                                         & \textcolor{SpringGreen}{\ding{55}} & \textcolor{SpringGreen}{\ding{55}} & \textcolor{SpringGreen}{\ding{55}} & \textcolor{Blue}{\ding{52}} & 5202       & 558   & 486   \\
$[ 96, 198 ]$ & $C_{2} \times C_{2} \times \text{SL}(2,3)$                   & \textcolor{SpringGreen}{\ding{55}} & \textcolor{SpringGreen}{\ding{55}} & \textcolor{SpringGreen}{\ding{55}} & \textcolor{SpringGreen}{\ding{55}} & 1347       & 224   & 198   \\
$[ 96, 199 ]$ & $Q_{8} \times A_{4}$                                         & \textcolor{SpringGreen}{\ding{55}} & \textcolor{SpringGreen}{\ding{55}} & \textcolor{SpringGreen}{\ding{55}} & \textcolor{Blue}{\ding{52}} & 5187       & 558   & 486   \\
$[ 96, 200 ]$ & $C_{2} \times (\text{SL}(2,3) \rtimes_\varphi C_{2})$        & \textcolor{Blue}{\ding{52}} & \textcolor{SpringGreen}{\ding{55}} & \textcolor{Blue}{\ding{52}} & \textcolor{SpringGreen}{\ding{55}} & 1332       & 224   & 198   \\
$[ 96, 201 ]$ & $(\text{SL}(2,3) \rtimes_\varphi C_{2}) \rtimes_\varphi C_{2}$ & \textcolor{SpringGreen}{\ding{55}} & \textcolor{SpringGreen}{\ding{55}} & \textcolor{SpringGreen}{\ding{55}} & \textcolor{SpringGreen}{\ding{55}} & 5202       & 558   & 486   \\
$[ 96, 202 ]$ & $(C_{2} \times \text{SL}(2,3)) \rtimes_\varphi C_{2}$        & \textcolor{SpringGreen}{\ding{55}} & \textcolor{SpringGreen}{\ding{55}} & \textcolor{SpringGreen}{\ding{55}} & \textcolor{SpringGreen}{\ding{55}} & 5202       & 558   & 486   \\
$[ 96, 203 ]$ & $(C_{2} \times C_{2} \times Q_{8}) \rtimes_\varphi C_{3}$    & \textcolor{SpringGreen}{\ding{55}} & \textcolor{SpringGreen}{\ding{55}} & \textcolor{SpringGreen}{\ding{55}} & \textcolor{Blue}{\ding{52}} & 189        & 0     & 0     \\
$[ 96, 204 ]$ & $((C_{2} \times D_{4}) \rtimes_\varphi C_{2}) \rtimes_\varphi C_{3}$ & \textcolor{SpringGreen}{\ding{55}} & \textcolor{SpringGreen}{\ding{55}} & \textcolor{SpringGreen}{\ding{55}} & \textcolor{Blue}{\ding{52}} & 189        & 0     & 0     \\
$[ 96, 226 ]$ & $C_{2} \times C_{2} \times S_{4}$                            & \textcolor{SpringGreen}{\ding{55}} & \textcolor{SpringGreen}{\ding{55}} & \textcolor{SpringGreen}{\ding{55}} & \textcolor{Blue}{\ding{52}} & 42         & 0     & 0     \\
$[ 96, 227 ]$ & $((C_{2} \times C_{2} \times C_{2} \times C_{2}) \rtimes_\varphi C_{3}) \rtimes_\varphi C_{2}$ & \textcolor{SpringGreen}{\ding{55}} & \textcolor{SpringGreen}{\ding{55}} & \textcolor{SpringGreen}{\ding{55}} & \textcolor{Blue}{\ding{52}} & 0          & 0     & 0     \\
$\Red{[ 96, 228 ]}$ & $C_{2} \times C_{2} \times C_{2} \times A_{4}$               & \textcolor{SpringGreen}{\ding{55}} & \textcolor{SpringGreen}{\ding{55}} & \textcolor{SpringGreen}{\ding{55}} & \textcolor{Blue}{\ding{52}} & 0          & 0     & 0     \\
$[ 96, 229 ]$ & $C_{2} \times ((C_{2} \times C_{2} \times C_{2} \times C_{2}) \rtimes_\varphi C_{3})$ & \textcolor{SpringGreen}{\ding{55}} & \textcolor{SpringGreen}{\ding{55}} & \textcolor{SpringGreen}{\ding{55}} & \textcolor{Blue}{\ding{52}} & 4779       & 853   & 720   \\

\end{longtable}
\end{center}

\subsection{Alternative Names for Some Small Groups}
\label{sec:smallgroups_alternative_names}

In \ref{tab:smallgroups_alternative_names} we list some alternative names for the groups that we have considered in this publication (cf.~\ref{tab:smallgroups_with_3dirrep}). To compile this list, we have made use of refs.~\cite{Miller:1916,fairbairn:1038,Bovier:1980gc,Fairbairn:1982jx,Altarelli:2010gt,Ishimori:2010au,Ludl:2009ft,Ludl:2010bj}.

\begin{center}
\footnotesize
\setcapindent{0em}
\rowcolors{1}{tableshade1}{white}
\renewcommand{\arraystretch}{1.4}
\newcolumntype{K}{>{\columncolor[gray]{.4}}c}
\newcolumntype{L}{>{\columncolor[gray]{.6}}c}
\newcolumntype{M}{>{\columncolor[gray]{.8}}c}
\newcolumntype{x}[1]{>{\centering\arraybackslash}p{#1}}
\begin{longtable}{|l|p{35ex}|c|c|}
\captionabove{Some aliases for the groups in \ref{tab:smallgroups_with_3dirrep}. The first and second columns give the \gapid{} and the group names displayed by \gap{}, respectively. The third column shows one or more alternative names that are in common use in the physics and mathematics literature. The fourth column, finally, gives a short description of the group where appropriate.}
\label{tab:smallgroups_alternative_names} 
\\

\hline \multicolumn{1}{|c|}{\gapid{}} & \multicolumn{1}{c|}{Group} & \multicolumn{1}{c|}{Other names} &\multicolumn{1}{c|}{Description} \\ 
\hline 
\endfirsthead

\multicolumn{4}{c}
{\tablename\ \thetable{} -- continued from previous page} \\
\hline \multicolumn{1}{|c|}{\gapid{}} & \multicolumn{1}{c|}{Group} & \multicolumn{1}{c|}{Other names} &\multicolumn{1}{c|}{Description} \\ 
\hline 
\endhead

\hline \multicolumn{4}{|r|}{{Continued on next page}} \\ \hline
\endfoot

\hline
\endlastfoot

$[ 12, 3 ] $ & $A_{4}$                                                 				               & $\Delta(12),T$ & Tetrahedral group    \\              
$[ 21, 1 ] $ & $C_{7} \rtimes_\varphi C_{3}$                             				       & $T_{7}$        &                        \\            
$[ 24, 3 ] $ & $\text{SL}(2,3)$                                           				       & $T'$               & Double cover of $A_4$\\          
$[ 24, 12 ]$ & $S_{4}$                                                   				       & $\Delta(24),O$ & Octahedral group           \\        
$[ 24, 13 ]$ & $C_{2} \times A_{4}$                                       				       & $\Sigma(24)$& Pyritohedral group          \\       
$[ 27, 3 ] $ & $(C_{3} \times C_{3}) \rtimes_\varphi C_{3}$              				       & $\Delta(27)$& 				\\	     
$[ 39, 1 ] $ & $C_{13} \rtimes_\varphi C_{3}$                            				       & $T_{13}$&				\\	     
$[ 42, 2 ] $ & $C_{2} \times (C_{7} \rtimes_\varphi C_{3})$                				       & $T_{14}$&				\\	     
$[ 48, 3 ] $ & $(C_{4} \times C_{4}) \rtimes_\varphi C_{3}$             			               & $\Delta(48)$&				\\	     
$[ 48, 28 ]$ & $\text{SL}(2,3) \rightarrow G \rightarrow C_{2}$            				       & &Double cover of $S_4$			\\	     
$[ 54, 8 ] $ & $((C_{3} \times C_{3}) \rtimes_\varphi C_{3}) \rtimes_\varphi C_{2}$                            & $\Delta(54)$&				\\	     
$[ 57, 1 ] $ & $C_{19} \rtimes_\varphi C_{3}$                              				       & $T_{19}$&				  \\           
$[ 60, 5 ] $ & $A_{5}$                                                     				       & $\Sigma(60),I$ & Icosahedral group	   \\          
$[ 63, 3 ] $ & $C_{3} \times (C_{7} \rtimes_\varphi C_{3})$                				       & $T_{21}$&				\\	     
$[ 75, 2 ] $ & $(C_{5} \times C_{5}) \rtimes_\varphi C_{3}$                                                    &$\Delta(75)$&				\\	     
$[ 78, 2 ] $ & $C_{2} \times (C_{13} \rtimes_\varphi C_{3})$                                                   &$T_{26}$ &				  \\           
$[ 81, 7 ] $ & $(C_{3} \times C_{3} \times C_{3}) \rtimes_\varphi C_{3}$                                       &$\Sigma(81)$ &				\\	     
$[ 84, 2 ] $ & $C_{4} \times (C_{7} \rtimes_\varphi C_{3})$                                                    &$T_{28}$ &				 \\            
$[ 93, 1 ] $ & $C_{31} \rtimes_\varphi C_{3}$                                                                  & $T_{31}$&			 	\\	     
$[ 96, 64 ]$ & $((C_{4} \times C_{4}) \rtimes_\varphi C_{3}) \rtimes_\varphi C_{2}$                            & $\Delta(96)$&				\\	     

\end{longtable}
\end{center}

\labelformat{section}{Appendix #1} 
\addtocontents{toc}{\protect\setcounter{tocdepth}{2}}

\clearpage
\newpage
\section{Construction of the Groups of Order at Most 100}
\label{sec:allgroups}

We will first describe how to generate all groups of order $\leq100$ in \gap{}. Then we will determine which groups have a 3-dimensional irrep and/or are a subgroup of \U{3} or \SU{3}. We include this information, because there seems to be a clear preference in model building for continuous or discrete subgroups of \U{3} or \SU{3}.

\subsection{Generating the Groups}
\label{sec:generate_groups}

The following lines of code generate the list of all groups of order $\leq100$ using the SmallGroups Library \cite{GAP4:smallgroups} in \gap{}:

\begin{Verbatim}[fontsize=\scriptsize,numbers=left,xleftmargin=20pt,formatcom=\color{gray}]
SizeScreen( [ 500, ] );
groups := AllSmallGroups([1..100]);;
for g in groups do
  Display(StructureDescription(g));
  Display(IdGroup(g));
  chartab := Irr(g);;
  for i in [1..Size(chartab)] do
    Print(chartab[i][1], " ");
    od;
    Print("\n");
  od;
time;
\end{Verbatim} 

These lines can be entered directly at the \gap{} prompt. In the following we assume that the preceding lines have been saved in a file named \code{smallgroups.gap} that is then loaded and automatically executed (see line 1 below):

\begin{Verbatim}[fontsize=\scriptsize,numbers=left,xleftmargin=20pt,formatcom=\color{gray}]
gap> Read("smallgroups.gap");
1
[ 1, 1 ]
1 
C2
[ 2, 1 ]
1 1 
C3
[ 3, 1 ]
1 1 1 
\end{Verbatim} 

We only display the first few lines of output (lines 2-10 above). For each group, there are 3 lines of output corresponding to lines 4, 5, 8 in the \gap{} code. For a non-trivial example, see lines 5-7 in the output. Line 5 displays the human-readable name of the group, line 6 gives its \gapid{} which uniquely identifies the group and which we will use as input for other \gap{} commands, and line 7 gives the first column of the character table, i.e.~the dimensions of the irreps \cite{Jones:1990ti}. 

\medskip

We find 1,048 groups of order $\leq100$ which we list in a separate file that we have made available for \href{http://lpsc.in2p3.fr/theorie/akin/documents/listof100smallgroups.pdf}{download} \cite{aw:2010:symmetries}. The first two columns of this list summarize the information we have obtained in this section.

\subsection[Groups that are Subgroups of $\SU{3}$ or $\U{3}$]{Groups that are Subgroups of $\boldsymbol{\SU{3}}$ or $\boldsymbol{\U{3}}$}
\label{sec:subgroups}

Next we determine which of these groups are subgroups of \U{3} or \SU{3}. If a group $\mathfrak{g}$ is (isomorphic to) a subgroup of \U{3}, there is a one-to-one correspondence between its elements and matrices of \U{3}. These matrices furnish a 3-dimensional {\it faithful} representation of $\mathfrak{g}$ that is not necessarily irreducible. Conversely, if $\mathfrak{g}$ has a 3-dimensional, faithful representation, then $\mathfrak{g}$ is a subgroup of \U{3}: For finite groups, every representation is equivalent to a unitary representation \cite{Jones:1990ti}, so the representation matrices are elements of \U{3}. By faithfulness, the representation $\rho$ is a one-to-one mapping between $\mathfrak{g}$ and the image of $\rho$ in \U{3}. By virtue of $\rho$ being a group homomorphism, $\operatorname{Im}\rho$ inherits the group properties from $\mathfrak{g}$, and consequently $\operatorname{Im}\rho\subset\U{3}$ is a group that is isomorphic to $\mathfrak{g}$. Finally, whether $\mathfrak{g}$ lies in \SU{3} can be verified by checking the determinant of representation matrices, since equivalent representations have the same determinant.

\medskip

The kernel of the representation is given by $\operatorname{Ker} \rho = \left\{g\in\mathfrak{g} | \operatorname{char}(g) =  \operatorname{char}(\mathds{1})\right\}$ \cite{Isaacs:2002}, and thus a 3-dimensional representation $\rho$ is faithful, iff $\mathds{1}$ is the only element whose character is $3$. For each of the 1,048 groups generated in \ref{sec:generate_groups} we calculate the character table. Below is the output for $A_4$:

\begin{Verbatim}[fontsize=\scriptsize,numbers=left,xleftmargin=20pt,formatcom=\color{gray}]
gap> g := SmallGroup(12,3);;
gap> Display(StructureDescription(g));
A4
gap> chartab := Irr(g);;
gap> Display(chartab);
[ [       1,       1,       1,       1 ],
  [       1,  E(3)^2,       1,    E(3) ],
  [       1,    E(3),       1,  E(3)^2 ],
  [       3,       0,      -1,       0 ] ]
\end{Verbatim} 

On line 1, we specify the group by entering its \gapid{} [12,3]. Lines 6-9 give its character table, where $\code{E(3)} = \exp{2\pi i/3}$ denotes the primitive third root of unity. The first column gives the dimensions of the representations: $\boldsymbol{1}, \boldsymbol{1'}, \boldsymbol{1''}, \boldsymbol{3}$. On line 9 corresponding to $\boldsymbol{3}$, there is only one character equal to 3, so $\boldsymbol{3}$ is faithful. This proves that $A_4$ is a subgroup of \U{3}. The representation matrices can be found by using the Repsn package \cite{GAP4:repsn} in \gap{}:

\begin{Verbatim}[fontsize=\scriptsize,numbers=left,xleftmargin=20pt,formatcom=\color{gray}]
gap> LoadPackage("repsn");;
gap> for i in [1..Size(chartab)] do                         
> Display(IrreducibleAffordingRepresentation(chartab[i]));
> od;
Pcgs([ f1, f2, f3 ]) -> [ [ [ 1 ] ], [ [ 1 ] ], [ [ 1 ] ] ]
Pcgs([ f1, f2, f3 ]) -> [ [ [ E(3)^2 ] ], [ [ 1 ] ], [ [ 1 ] ] ]
Pcgs([ f1, f2, f3 ]) -> [ [ [ E(3) ] ], [ [ 1 ] ], [ [ 1 ] ] ]
Pcgs([ f1, f2, f3 ]) -> [ [ [ 0, 1, 0 ], [ 0, 0, 1 ], [ 1, 0, 0 ] ], 
  [ [ -1, 0, 0 ], [ 0, 1, 0 ], [ 0, 0, -1 ] ], 
  [ [ -1, 0, 0 ], [ 0, -1, 0 ], [ 0, 0, 1 ] ] ]
\end{Verbatim} 

Lines 8,9,10, respectively, correspond to the representation matrices of the generators \code{f1}, \code{f2}, \code{f3} of $A_4$ for the 3-dimensional irrep. Their determinants are all 1, and thus $A_4$ is a subgroup of \SU{3}.

\medskip

In other cases, when there is no faithful, irreducible 3-dimensional representation, we have to consider the reducible ones. If $A, B$ are two representations, then $\operatorname{char}(A\oplus B) = \operatorname{char}(A) + \operatorname{char}(B)$, i.e.~we obtain the character of $A\oplus B$ by adding the rows in the character table that correspond to $A$ and $B$. For a given group, we consider all direct sums that are 3-dimensional and calculate their characters. For each direct sum, the first element of the character will be 3, corresponding to the dimension of the representation. If there is more than one 3, we conclude that the direct sum is not faithful. If we cannot find any direct sum that is faithful, we conclude that $\mathfrak{g}$ is not isomorphic to a subgroup of \U{3}.

\medskip

Assume that we can find a 3-dimensional faithful, reducible representation, thereby proving that $\mathfrak{g}$ is a subgroup of \U{3}. The representation matrices are block-diagonal, and each submatrix is unitary. There are two cases: All submatrices are $1\times1$, or one is $2\times2$ and the other is $1\times1$. In the former case, the representation matrices are diagonal and commute, thus $\mathfrak{g} \simeq \mathbb{Z}_p\times\mathbb{Z}_q\times\mathbb{Z}_r \subset \U{1}^n$ for some $n\leq3$. In the latter case, we consider the canonical embedding of the submatrices into \U{3} (i.e.~by extending the submatrix by the identity matrix to match the dimensions). Every representation matrix can be uniquely written as a product of these embedded submatrices, and the submatrices corresponding to different blocks trivially commute. This establishes that $\mathfrak{g}$ is isomorphic to a subgroup of $\U{2}\times\U{1}$.

\bigskip

We have implemented the above algorithm in a \gap{} script. The results have been summarized in ref.~\cite{aw:2010:symmetries} and made available for \href{http://lpsc.in2p3.fr/theorie/akin/documents/listof100smallgroups.pdf}{download}.

\subsection{Comparing Our Results to the Existing Literature}

We have compared our results to the existing literature on \SU{3} subgroups \cite{Miller:1916,fairbairn:1038,Bovier:1980gc,Fairbairn:1982jx,Ludl:2009ft,Ludl:2010bj}. Identifying the groups is not always straightforward, since they may appear under different names in different contexts, e.g.~$A_4$ is listed as $\Delta(12)$ in ref.~\cite{fairbairn:1038} and as part of the $C$ series in ref.~\cite{Miller:1916}.

In \href{http://lpsc.in2p3.fr/theorie/akin/documents/listof100smallgroups.pdf}{Tab.~1} of ref.~\cite{aw:2010:symmetries} we list all groups of order at most 100 and for each group indicate whether it is a subgroup of \U{3}, \SU{3}, \U{2}, \SU{2}, $\U{2}\times\U{1}$, $\SU{2}\times\U{1}$, respectively.

We find that the groups in our list that are subgroups of \SU{3} but not of $\U{2}\times\U{1}$ agree with those in ref.~\cite{fairbairn:1038,Bovier:1980gc,Fairbairn:1982jx} except in the following cases: According to ref.~\cite{Fairbairn:1982jx}, the groups \gid{42}{2}, \gid{78}{2}, \gid{84}{2} are in \SU{3}, but our analysis along the lines of \ref{sec:subgroups} shows that they are only in \U{3}, in \SU{3}. 

Ref.~\cite{Ludl:2010bj} only explicitly lists groups that are not direct products with cyclic factors, and thus does not consider the groups in question, but according to Theorem II.2 in the same publication, these groups are in \U{3}, so we have agreement.

Also, the groups \gid{36}{11}, \gid{72}{42}, \gid{81}{9} and \gid{84}{11} were not listed in refs.~\cite{fairbairn:1038,Bovier:1980gc,Fairbairn:1982jx}, but we have verified that these groups are indeed in \SU{3}. 
\gid{81}{9} and \gid{84}{11} are part of the C series as given in ref.~\cite{Miller:1916}. This has already been pointed out by ref.~\cite{Ludl:2010bj}. The Groups \gid{36}{11} and \gid{72}{42} have not been explicitly listed in ref.~\cite{Ludl:2010bj}, but the discussion following Theorem II.2 in the same publication makes it clear that these groups are in \SU{3}.
  
Ref.~\cite{Ludl:2010bj} in Section II.1 makes the observation that a finite subgroup of \U{3} is not in \U{2} or \U{1}, if and only if it has a faithful 3-dimensional irreducible representation. In our analysis, we find counterexamples: E.g.~the group $\gid{16}{3}\subset \U{3}$ is not in \U{2} and has no 3-dimensional irreducible representation. The reason is that the existence of a \textit{reducible and faithful} 3-dimensional representation is already sufficient for being a subgroup of \U{3}. For more details, see \ref{sec:subgroups}.

For generating the $\Delta(3n^2)$ series, we have used the generators from Tab.~1 in ref.~\cite{fairbairn:1038} with $j=1$ and $k=0$ (also see ref.~\cite{Luhn:2007uq}). Note that if we take some arbitrary integers $j$ and $k$, the representation may not be faithful and thus will not generate (a subgroup of \U{3} that is isomorphic to) $\Delta(3n^2)$. Also note that in ref.~\cite{fairbairn:1038} the generators for $\Sigma(360)$ generate a group of order 1,080 which is a non-split extension\footnote{We are indebted to Patrick Otto Ludl for pointing out that it is not the direct product of $A_6$ and $C_3$ as we had incorrectly identified in the first version of this publication due to a misinterpretation of the \gap{} output.} of $A_6$ by $C_3$, and not $A_6$. We agree with ref.~\cite{Miller:1916} that lists the same group as $\Sigma(360 \phi)$.

\subsection[Groups that Contain $A_4$ as a Subgroup]{Groups that Contain \bs{A_4} as a Subgroup}

Since many publications in the past have highlighted $A_4$ and its connection to tribimaximal mixing, we find it useful to list the groups that contain $A_4$ as a subgroup:

\begin{Verbatim}[fontsize=\scriptsize,numbers=left,xleftmargin=20pt,formatcom=\color{gray}]
LoadPackage("sonata");
for n in [1..100] do
  for g in AllSmallGroups(n) do
    sg := Subgroups(g);
    if "A4" in List(sg,x->StructureDescription(x)) then
      Print(IdGroup(g),"\n");
    fi;
  od;
  UnloadSmallGroupsData();
od;
\end{Verbatim} 

For every $n$ from 1 to 100 (line 2), we generate all groups of order $n$ (line 3). For each such group, we determine its subgroups (line 4) and check whether $A_4$ is one of them (line 5). If the answer is positive, we print the \gapid{} of the respective group (line 6). On a technical note, since the number of subgroups becomes large, we need to increase the default memory allocation for \gap{}. The results are presented in the last column of \href{http://lpsc.in2p3.fr/theorie/akin/documents/listof100smallgroups.pdf}{Tab.~1} in ref.~\cite{aw:2010:symmetries}.

\clearpage
\newpage
\section{Breaking the Family Symmetry to Subgroups}
\label{sec:breaking_the_family_symmetry}

In general, the \vev{} of a flavon field in an $n$-dimensional representation can take any value in $\mathbb{R}^n$, but since a finite group has only finitely many subgroups, there will be a finite number of \textit{inequivalent} \vevs{} that break to different subgroups. It should be noted that the neutrino mixing angles will in general depend on the \textit{length} of the \vevs{}. Rescaling all \vevs{} with the same factor, though, does not have any effect on the mixing angles.

\medskip

For definiteness, we again choose $A_4\times C_3$ as a working example, but it should be clear that the following discussion is completely general. Consider the \gap{} script listed below:
\begin{Verbatim}[fontsize=\scriptsize,numbers=left,xleftmargin=20pt,formatcom=\color{gray}]
SizeScreen( [ 200, ] );
LoadPackage("repsn");
LoadPackage("sonata");
g := SmallGroup(36,11);
chi:=Irr(g);
sg:=Subgroups(g);
Display(Size(sg));
Display(List(sg, StructureDescription));
for n in [1..Size(chi)] do
  Print("Representation"," ",n,"\n\n");
  rep:=IrreducibleAffordingRepresentation(chi[n]);
  for a in sg do
    a1:=List(a,x->TransposedMat(x^rep));
    Print(StructureDescription(a),"\t\t",BaseFixedSpace(a1),"\n");
  od;
Print("\n\n\n");
od;
\end{Verbatim}
\label{verb:symbreak}
Line 4 sets the group to $\mathfrak{g} = A_4\times C_3$, and lines 5 and 6 calculate the character table and the subgroups, respectively. Line 7 prints the number of the subgroups (30 in this case), and line 8 lists each subgroup by its human-readable name. 

Lines 9-17 loop over the 12 irreducible representations $\rho_i$ of $\mathfrak{g}$. For each representation, line 11 generates the representation matrices, and lines 12-15 loop over the 30 subgroups $\mathfrak{h}_j$ ($j=1,\ldots,30$) to find the \vev{} that breaks $\mathfrak{g}$ to $\mathfrak{h}_j$ as follows: For each $\mathfrak{h}_j$, line 13 generates the list of the matrices $M_k$ ($k=1,\ldots,\text{dim}\,\mathfrak{h}_j$) that correspond to the representation $\rho_i$ and elements in $\mathfrak{h}_j$. Line 14 calculates a basis of the common eigenspace of the $M_k$ for the eigenvalue 1, i.e.~it calculates the \vevs{} that are left invariant by the action of $\mathfrak{h}_j$. In the following, we will call this eigenspace $V_{\mathfrak{h}_j}$.

From the construction it is clear that the obtained \vevs{} leave $\mathfrak{h}_j$ intact, but are not guaranteed to break \textit{exactly} to $\mathfrak{h}_j$, i.e.~there may be another subgroup under which the \vevs{} do not transform. In that case, one of the subgroups is contained in the other, or both are contained in a third subgroup of $\mathfrak{g}$. Thus, if more than one $\mathfrak{h}_j$ leads to the same \vev{}(s), the largest one is the unbroken symmetry group. This establishes a correspondence between the subgroups of $\mathfrak{g}$ and the vector spaces of \vevs{} that break to them, i.e.~$\mathfrak{h}_j \leftrightarrow V_{\mathfrak{h}_j}$ (where $\mathfrak{h}_j$ is the \textit{largest} subgroup to which $V_{\mathfrak{h}_j}$ breaks). If there are no \vevs{} that break to $\mathfrak{h}_j$, then $V_{\mathfrak{h}_j}$ is the empty set. Note that subgroups that are conjugate to each other correspond to different embeddings in $\mathfrak{g}$ and lead to different physics, so for our purposes they are not equivalent. To contribute to the clarity of the current discussion, we present the results for the irrep \bs{3} in \vref{tab:invariant_subspaces}. Note that $V^{\rho_{10}}_{\mathfrak{h}_{22}}$ in \ref{tab:invariant_subspaces} corresponds to the \vev{} of the flavon field $\varphi_T$ in \vref{tab:altferug-symbreak} (The flavon fields $\varphi_s$ and $\xi$ transform as \bs{3'} and \bs{1''}, respectively, and the corresponding \vev{} spaces can easily be determined by the same \gap{} script.)

\medskip

\begin{table}[h!]
\begin{center}
\renewcommand{\arraystretch}{1.2}
\begin{tabular}{|l|l|}
\hline
  $\mathfrak{h}_{5\phantom{2}} = C_3$ & $V^{\rho_{10}}_{\mathfrak{h}_5} = \mathrm{span}\,\left\{( 1, 0, 0 ),\, ( 0, 1, 0 ),\, ( 0, 0, 1 )\right\}$ \\
  $\mathfrak{h}_{19} = C_6$ & $V^{\rho_{10}}_{\mathfrak{h}_{19}} = \mathrm{span}\,\left\{( 0, 1, 0 )\right\}$ \\
  $\mathfrak{h}_{20} = C_6$ & $V^{\rho_{10}}_{\mathfrak{h}_{20}} = \mathrm{span}\,\left\{( 0, 0, 1 )\right\}$ \\
  $\mathfrak{h}_{21} = C_6$ & $V^{\rho_{10}}_{\mathfrak{h}_{21}} = \mathrm{span}\,\left\{( 1, 0, 0 )\right\}$ \\
  $\mathfrak{h}_{22} = C_3 \times C_3$ & $V^{\rho_{10}}_{\mathfrak{h}_{22}} = \mathrm{span}\,\left\{( 1, 1, 1 )\right\}$ \\
  $\mathfrak{h}_{23} = C_3 \times C_3$ & $V^{\rho_{10}}_{\mathfrak{h}_{23}} = \mathrm{span}\,\left\{( \m1, \m1, 1 )\right\}$ \\
  $\mathfrak{h}_{24} = C_3 \times C_3$ & $V^{\rho_{10}}_{\mathfrak{h}_{24}} = \mathrm{span}\,\left\{(  \m1, 1, 1 )\right\}$ \\
  $\mathfrak{h}_{25} = C_3 \times C_3$ & $V^{\rho_{10}}_{\mathfrak{h}_{25}} = \mathrm{span}\,\left\{( 1, \m1, 1 )\right\}$\\
  $\mathfrak{h}_{30} = A_4 \times C_3$ & $V^{\rho_{10}}_{\mathfrak{h}_{30}} = \mathrm{span}\,\left\{( 0, 0, 0 )\right\}$\\
\hline
\end{tabular}
\end{center}
\setcapindent{0em}
\caption{The common eigenspaces for the eigenvalue 1 of the representation matrices for the irrep \bs{3} that leave the respective subgroup $\mathfrak{h}_j$ invariant. The irrep \bs{3} corresponds to the 10th line of the character table, hence its name $\rho_{10}$. The numbering of the subgroups $\mathfrak{h}_j$ corresponds to their order given by the \gap{} script on p.~\pageref{verb:symbreak}. If an $\mathfrak{h}_j$ is not listed here, this means that there exist no \vevs{} for the irrep \bs{3} that can break to this particular subgroup.}
\label{tab:invariant_subspaces}
\end{table}

Inspecting \vref{tab:invariant_subspaces}, we see another subtlety that we have to take into account (for clarity, we drop the superscripts). $V_{\mathfrak{h}_5}$ is spanned by three \vevs{}, and varying them independently breaks $\mathfrak{g}\rightarrow\mathfrak{h}_5$. On the other hand, $V_{\mathfrak{h}_{19}}$ is a subset of $V_{\mathfrak{h}_5}$ and breaks to a larger symmetry $\mathfrak{g}\rightarrow\mathfrak{h}_{19}$. The same is true for $V_{\mathfrak{h}_{20}}$, \ldots, $V_{\mathfrak{h}_{25}}  \subset V_{\mathfrak{h}_5}$ that break to $\mathfrak{h}_{20}$, \ldots, $\mathfrak{h}_{25} \supset \mathfrak{h}_{5}$, respectively. 

Assume that we choose a \vev{} $v$ that lies in $V_{\mathfrak{h}_{5}}$, but in none of the $V_{\mathfrak{h}_{19}}$, \ldots, $V_{\mathfrak{h}_{25}}$. Then $\mathfrak{g}$ will necessarily break to $\mathfrak{h}_{5}$, since neither of the larger symmetries leave $v$ invariant. Thus, we can effectively break $\mathfrak{g}\rightarrow\mathfrak{h}_5$ with a single \vev{}, although $V_{\mathfrak{h}_{5}}$ is 3-dimensional. In general, a \vev{} will break to the symmetry that corresponds to the \textit{smallest} vector space $V_{\mathfrak{h}_{j}}$ in which it is contained. For two vector spaces $A$, $B$ we \textit{define} $A \leq B$, if $A\subset B$ (this defines a partial ordering). A partially ordered set does not necessarily have a smallest element, but in this case it does. Given a \vev{} $v$, there is always a smallest $V_{\mathfrak{h}_{j}}$ that contains it: Suppose $v \in V_{\mathfrak{h}_{i_0}}\cap V_{\mathfrak{h}_{i_1}}$ and $V_{\mathfrak{h}_{i_0}}\not\subset V_{\mathfrak{h}_{i_1}}$ and $V_{\mathfrak{h}_{i_1}}\not\subset V_{\mathfrak{h}_{i_0}}$. Then $\mathfrak{h}_{i_0} \neq \mathfrak{h}_{i_1}$ and $v$ will break to a group $\mathfrak{h}_{i_2}$ that contains both $\mathfrak{h}_{i_0}$ and $\mathfrak{h}_{i_1}$ as proper subgroups, and as a consequence $V_{\mathfrak{h}_{i_2}}\subsetneq V_{\mathfrak{h}_{i_0}}$ and $V_{\mathfrak{h}_{i_2}}\subsetneq V_{\mathfrak{h}_{i_1}}$. So either there is a smaller vector space $V_{\mathfrak{h}_{i_2}}$ in which $v$ lies or the assumptions are not correct, i.e.~$V_{\mathfrak{h}_{i_0}}\cap V_{\mathfrak{h}_{i_1}} = \emptyset$ or $V_{\mathfrak{h}_{i_0}} \subset V_{\mathfrak{h}_{i_1}}$ or $V_{\mathfrak{h}_{i_1}} \subset V_{\mathfrak{h}_{i_0}}$. Thus, $v$ is in one and only one smallest vector space.

On a practical note, to break $\mathfrak{g}\rightarrow\mathfrak{h}$, one must pick a \vev{} $v \in V_{\mathfrak{h}}$ such that $v \not\in V_{\mathfrak{m}}$ for all $V_{\mathfrak{m}} \subset V_{\mathfrak{h}}$. In the following, we will always make this assumption without explicitly stating it.

\medskip

In the present publication, we consider models with up to 3 flavon fields and they do not necessarily transform in the same representation, so in the following we extend the present analysis to this more general case.

\medskip

\begin{algorithm}
\SetAlgoLined
Choose irreps $\rho_{i_0}$, $\rho_{i_1}$, $\rho_{i_2}$ with $i_0, i_1, i_2 \in \left\{1,\ldots,12\right\}$\;
For $\rho_{i_0}$ find \vev{} space $V^{i_0}_{\mathfrak{h}_k}$ that breaks $\mathfrak{g} \longrightarrow \mathfrak{h}_k$ for $k \in \left\{1,\ldots,30\right\}$\;\nllabel{nl:i0}
For $\rho_{i_1}$ find \vev{} space $V^{i_1}_{\mathfrak{h}_k}$ that breaks $\mathfrak{g} \longrightarrow \mathfrak{h}_k$ for $k \in \left\{1,\ldots,30\right\}$\;\nllabel{nl:i1}
For $\rho_{i_2}$ find \vev{} space $V^{i_2}_{\mathfrak{h}_k}$ that breaks $\mathfrak{g} \longrightarrow \mathfrak{h}_k$ for $k \in \left\{1,\ldots,30\right\}$\;\nllabel{nl:i2}
\For{$p$ \KwSty{from} $1$ \KwTo $30$}{\nllabel{nl:begfor}
\lIf{$V^{i_0}_{\mathfrak{h}_p} = \emptyset$}{continue with next $p$-iteration\;}
\For{$q$ \KwSty{from} $1$ \KwTo $30$}{
\lIf{$V^{i_1}_{\mathfrak{h}_q} = \emptyset$}{continue with next $q$-iteration\;}
\For{$r$ \KwSty{from} $1$ \KwTo $30$}{
\lIf{$V^{i_2}_{\mathfrak{h}_r} = \emptyset$}{continue with next $r$-iteration\;}
$\mathfrak{t} = \mathfrak{h}_p \cap \mathfrak{h}_q \cap \mathfrak{h}_r$\;\nllabel{nl:intersection}
$\mathfrak{T} = \mathfrak{T} \cup \left\{ \left( \mathfrak{t}, V^{i_0}_{\mathfrak{h}_p}, V^{i_1}_{\mathfrak{h}_q}, V^{i_2}_{\mathfrak{h}_r}\right)\right\}$\;\nllabel{nl:save}
}
}
}\nllabel{nl:endfor}
\caption{How to find the \vevs{} that break to a specific subgroup.}
\label{alg:finding_vevs}
\end{algorithm}

\medskip

In Algorithm \vref{alg:finding_vevs}, we describe our approach to finding all inequivalent \vevs{} that break to different subgroups in the case of three flavon fields. First, we choose arbitrary, but fixed representations $\rho_{i_0}$, $\rho_{i_1}$, $\rho_{i_2}$ for those fields. Next, for each representation separately, we apply the above procedure for finding the \vevs{} that break to all subgroups (lines \ref{nl:i0}-\ref{nl:i2}). When all \vevs{} are turned on, the unbroken symmetry is the intersection of $\mathfrak{h}_p$, $\mathfrak{h}_q$, $\mathfrak{h}_r$ corresponding to the \vevs{} $V^{i_0}_{\mathfrak{h}_p}, V^{i_1}_{\mathfrak{h}_q}, V^{i_2}_{\mathfrak{h}_r}$, respectively. Thus, for each representation $\rho_{i_0}$, $\rho_{i_1}$, $\rho_{i_2}$, we loop over the 30 subgroups (lines \ref{nl:begfor}-\ref{nl:endfor}), calculate the intersections (line \ref{nl:intersection}) and collect the breaking patterns (line \ref{nl:save}). We consider two breaking patterns as different, if the respective \vevs{} $V^{i_0}_{\mathfrak{h}_p}, V^{i_1}_{\mathfrak{h}_q}, V^{i_2}_{\mathfrak{h}_r}$ do not coincide (independent of the subgroups $\mathfrak{t}$ coinciding or not).

\medskip

One may wonder whether we get all the possible breaking patterns with three flavon fields, and the answer is positive: A single \vev{} will necessarily breaks to a group (for a finite group, it is sufficient to show closure; if some elements of $\mathfrak{g}$ leave the \vev{} invariant and are thus part of the unbroken symmetry, then this will also be true for any product of those elements; hence the set is closed under the group multiplication). Since $\mathfrak{g}$ is finite, we can enumerate all its subgroups and find \vevs{} that break to them (that is what we did in the \gap{} script on p.~\pageref{verb:symbreak}), thus establishing a correspondence $\mathfrak{h}_i \leftrightarrow V_{\mathfrak{h}_i}$ between subgroups and \vevs{} (for simplicity, we have dropped the superscripts denoting the dependence on the representation). 

Suppose we are given two \vevs{} $v_1$, $v_2$ that break $\mathfrak{g} \rightarrow \mathfrak{h}$. Individually, $v_1$, $v_2$ will break to some $\mathfrak{h}_1$, $\mathfrak{h}_2$, respectively, and $\mathfrak{h} = \mathfrak{h}_1 \cap \mathfrak{h}_2$. Note that there may be more than one pair $\mathfrak{h}_1$, $\mathfrak{h}_2$ whose intersection is $\mathfrak{h}$, and that is why we cannot consider breaking patterns as equivalent that lead to the same $\mathfrak{t}$ in line \ref{nl:save} of Algorithm \vref{alg:finding_vevs}. The earlier established correspondence between subgroups and \vevs{} now gives us $\mathfrak{h}_1 \leftrightarrow V_{\mathfrak{h}_1}$ and $\mathfrak{h}_2 \leftrightarrow V_{\mathfrak{h}_2}$, and we can conclude that $v_1 \in V_{\mathfrak{h}_1}$ and  $v_2 \in V_{\mathfrak{h}_2}$. Thus, the iteration in Algorithm \vref{alg:finding_vevs} includes the symmetry breaking pattern $\mathfrak{g} \rightarrow \mathfrak{h}$ for any two \vevs{} $v_1$ and $v_2$. This concludes the proof for two \vevs{}, and the generalization to the case of three \vevs{} is straightforward.

\medskip

In this section, we have solved the problem of finding the \vevs{} that break to all subgroups of a given symmetry in full detail and generality. It is important to note, though, that the neutrino mixing angles will not only depend on the residual family symmetry $\mathfrak{h}$, but also on the alignment and magnitude of the \vevs{} in $V_{\mathfrak{h}}$ (as usual we assume that $v$ does not lie in a smaller subspace). Resolving this ambiguity is beyond the scope of group theory and needs to be addressed by model building.

\clearpage
\newpage
\section{Clebsch-Gordan Coefficients for Finite Groups}
\label{app:clebsch-gordan-coefficients}

Currently, it is general practice to construct the Clebsch-Gordan coefficients (CGCs) for the various groups that are studied in physics on a case-by-case basis using heuristic methods. It is clear that such an approach becomes cumbersome, if one considers more than one group or the number of irreducible representations is large. Also, for automating the steps from the choice of the family symmetry to finding the invariant Lagrangian to calculating the mixing angles and phases, we need a systematic way of deriving the CGCs that does not rely on the specifics of the group under consideration.

\medskip

An algorithm due to P.~M.~van den Broek and J.~F.~Cornwell \cite{vandenbroek:1978aa} solves this problem in full generality: Given the character table and the explicit form of the unitary representation matrices, it calculates the CGCs for any finite group. We have implemented this algorithm in \python{} to automatically generate the CGCs for any finite group. We get the character table and the representation matrices from \gap{} that we have interfaced with our \python{} programs to achieve a high level of automization.

\medskip

To establish our notation for the CGCs and to contribute to the clarity of the discussion in \ref{sec:altarelli-feruglio-model}, we present an explicit example of how to contract the indices in the tensor product of any 2 fields transforming in irreducible representations of the family symmetry. Consider e.g.~2 fields $\phi$ and $\psi$ that transform as \bs{3}'s of $A_4\times C_3$. The product $\phi_i\,\psi_j$ with $i,j=1,2,3$ transforms as

\begin{equation}
\bs{3} \times \bs{3} = \bs{1}  +  \bs{1'''}  +  \bsn{1}{4}  +  2 \times \bs{3}
\label{eq:3x3tensorproduct}
\end{equation}
 
and contains the singlet representation \bs{1}, i.e.~for some choice of $i,j$, the product does not transform. The CGCs give the change of basis between the right-hand side and left-hand side of \ref{eq:3x3tensorproduct}:
\begin{equation}
C_{ijk} \, \bs{3}_i \, \bs{3}_j = \bs{1}_k, \quad k=1, \quad i,j=1,2,3
\label{eq:CGC3x3equal1}
\end{equation}
In the general case $\bs{p}\otimes\bs{q}=\sum n_{pq}^{r} \times \bs{r}$, the CGCs will depend on the representations \bs{p}, \bs{q}, \bs{r} and on the number of times $n_{pq}^{r}$ that \bs{r} appears in the decomposition of the tensor product. For the case of \ref{eq:CGC3x3equal1}, our \python{} script gives:

\begin{equation}
C(p=\bs{3},q=\bs{3},r=\bs{1},n_{pq}^{r}=1)_{ij;\,k=1} = \left(\begin{matrix}
\frac{1}{\sqrt{3}} & 0  & 0 \\
0 & \frac{1}{\sqrt{3}} & 0 \\
0 & 0 & \frac{1}{\sqrt{3}} \\
\end{matrix}\right)
\end{equation}

Since \bs{1} is a singlet, the index $k$ only takes one value so that we can write the CGCs in the form of a matrix where $i,j$ label the rows and columns, respectively. Coming back to the case of our 2 fields $\phi$ and $\psi$, we conclude that the combination

\begin{equation}
\frac{1}{\sqrt{3}}\phi_1 \psi_1 + \frac{1}{\sqrt{3}}\phi_2 \psi_2 + \frac{1}{\sqrt{3}}\phi_3 \psi_3 
\label{eq:exampleforsinglet}
\end{equation}

is invariant under $A_4\times C_3$. Note that the CGCs depend on the choice of the representation matrices. One can check the invariance of \ref{eq:exampleforsinglet} by using the explicit form of the representation matrices obtained from the \gap{} script that we have discussed on p.~\pageref{verb:gap_group_properties}.

\medskip

We performed several checks to ascertain that the CGCs are calculated correctly. For one thing, we have compared our output to the (comparatively few) results that exist in the literature. For another, we have checked for all 90 groups in \vref{tab:smallgroups_with_3dirrep} and for all irreducible representations $\bs{p}$, $\bs{q}$ that the terms $C_{ijk} \, \bs{p}_i \, \bs{q}_j$ transform as $\bs{r}_k$ where $\bs{p}\otimes\bs{q}=\sum n_{pq}^{r} \times \bs{r}$.

We find complete agreement except for $A_5$. The problem can be traced back to the fact that the representation matrices for $A_5$ provided by \gap{} are not unitary. After choosing unitary representation matrices, the algorithm gives the correct CGCs also for this remaining case.

\clearpage
\newpage
\section{Elements of Finite Group Theory}
\label{app:elements_of_finite_group_theory}

In this appendix we summarize some of the most important definitions and theorems from finite group theory that we use in the present publication. 

\subsection{Direct Products}

Given 2 groups $A$, $B$ we define their \textit{direct product} $A\times B$ as the set of all pairs $(a,b)$ with $a\in A$ and $b\in B$, where the group operation is defined by element-wise multiplication:
\begin{equation}
(a_1,b_1)\cdot(a_2,b_2) \equiv (a_1\cdot a_2,b_1\cdot b_2)
\label{eq:direct_product_multiplication_rule}
\end{equation}
Conversely, when a group $G$ is given, we can ask whether we can write it as the direct product of 2 of its subgroups, say $A$ and $B$. From our previous definition it is clear that we would like 2 conditions to be fulfilled: Firstly, every element $g\in G$ should be expressible as a product $g=a\cdot b$ with $a\in A$ and $b\in B$. If $A$ and $B$ have no common elements except for the identity element in $G$, it easily follows from the group properties that this decomposition is unique, and as such, we have a one-to-one correspondence $a\cdot b\leftrightarrow(a,b)$. Secondly, all elements in $A$ should commute with all elements in $B$ so that we can mimic the product in \ref{eq:direct_product_multiplication_rule}:
\begin{equation}
g_1 \cdot g_2 = (a_1 \cdot b_1) \cdot (a_2 \cdot b_2) = a_1 \cdot b_1 \cdot a_2 \cdot b_2 = a_1 \cdot a_2 \cdot b_1 \cdot b_2 = (a_1 \cdot a_2) \cdot (b_1 \cdot b_2)
\label{eq:direct_product_definition_of_multiplication}
\end{equation}
With the identification $a\cdot b\leftrightarrow(a,b)$, the previous line reads:
\begin{equation}
g_1\cdot g_2 = (a_1,b_1)\cdot(a_2,b_2) = (a_1\cdot a_2,b_1\cdot b_2)
\end{equation}
Thus we have proven that there is a one-to-one correspondence $a\cdot b\leftrightarrow(a,b)$ between the elements of $G$ and $A\times B$ that is compatible with the group operation, i.e.~the 2 groups are \textit{isomorphic}. For all practical purposes, we can view these 2 groups to be identical and write $G=A\times B$.

\subsection{Normal Subgroups}

A subgroup $N\subset G$ is called \textit{normal}, if for any $n\in N$ it holds that $g n g^{-1}\in N$ for all $g\in G$, i.e.~the operation of conjugation with an arbitrary element of $G$ maps $N$ into itself. One then writes $N\vartriangleleft G$. This concept is relevant in the present context, because both $A$ and $B$ are normal subgroups of $G=A\times B$:
\begin{equation}
g a' g^{-1} = (ab) a' (ab)^{-1} = a b a' b^{-1} a^{-1} = a b b^{-1} a' a^{-1} = a a' a^{-1} \in A
\end{equation}
The proof for $B$ is analogous. If $N$ is normal in $G$, the cosets of $G$ with respect to $N$ form a group, called the quotient group and denoted by $G/N$. Dividing by $N$ means that we identify all elements that differ by multiplication by $n\in N$, i.e.~$g_1 \sim g_2$, iff $g_1=n g_2$. In this sense, it is clear that $(A\times B) / B$ is the set of all elements of the form $(a,\bs{1})$, which is isomorphic to $A$, because $(a,\bs{1})\sim a\cdot \bs{1} = a$. An analogous statement holds for $B$.

\subsection{Semidirect Products}

The semidirect product is a straightforward generalization of the direct product for the case that $A$ and $B$ do not commute. Let us repeat the calculation in \ref{eq:direct_product_definition_of_multiplication}:
\begin{equation}
g_1 \cdot g_2 = (a_1 \cdot b_1) \cdot (a_2 \cdot b_2) = a_1 \cdot b_1 \cdot a_2 \cdot b_2 = a_1 \cdot a_2 \cdot a_2^{-1} \cdot b_1 \cdot a_2 \cdot b_2 = a_1 \cdot a_2 \cdot (a_2^{-1} \cdot b_1 \cdot a_2) \cdot b_2 
\label{eq:semidirect_product_definition_of_multiplication}
\end{equation}
Firstly, if we want to have any chance of writing $g_1\cdot g_2$ as a product $\tilde{a}\cdot\tilde{b}$ with $\tilde{a}\in A$ and $\tilde{b}\in B$, we have to assume that $a_2^{-1} \cdot b_1 \cdot a_2 \in B$. Since this must hold for all $a_2\in A$ (and trivially holds for all $b\in B$), this is equivalent to requiring that $B$ be a normal subgroup. Secondly, we need to know the action of $A$ on $B$ by conjugation which we denote by $\varphi_a$:
\begin{equation}
\varphi_a: B \rightarrow B, \qquad \varphi_a(b) = a^{-1} \cdot b \cdot a
\label{eq:definition_of_action_of_A_on_B}
\end{equation}
Note that $\varphi_a: b\mapsto a^{-1}\cdot b\cdot a$ is an automorphism of $B$, and $\varphi: a \mapsto \varphi_a$ is a homomorphism from $A$ to $\mathrm{Aut}(B)$, the automorphism group of $B$. This may look like notational overkill, but it will become clear in a moment why we chose to do so. Now we can write
\begin{equation}
g_1\cdot g_2 = (a_1,b_1)\cdot(a_2,b_2) = (a_1\cdot a_2,\, \varphi_{a_2}(b_1) \cdot b_2),
\label{eq:definition_of_a_semidirect_product}
\end{equation}
in analogy to \ref{eq:direct_product_multiplication_rule}, and the only difference is that the rule for multiplication gets slightly modified. If $A$ and $B$ are given as subgroups of $G$, the multiplication rule between elements from $A$ and $B$ is known, and we can calculate the right-hand side of the second term in \ref{eq:definition_of_action_of_A_on_B}. If, however, we are given 2 groups $A$ and $B$ that bear no relation to each other, we have to \textit{choose} $\varphi_a$ from the set of automorphisms of $B$, and this \textit{serves as the definition} of conjugation. In this sense, the semidirect product is not unique, since it depends on the choice of $\varphi$.

\medskip

To summarize, the \textit{semidirect product} of $A$ and $B$ with respect to $\varphi$ is the set of all pairs $(a,b)$ with $a\in A$ and $b\in B$, where the group operation is defined by \ref{eq:definition_of_a_semidirect_product} and $\varphi: a \mapsto \varphi_a$ is a homomorphism from $A$ to $\mathrm{Aut}(B)$. We use the notation by $G \equiv B \rtimes_\varphi A$ for the semidirect product. Then, \begin{inparaenum}[(i)] \item $B\vartriangleleft G$, i.e.~$B$ is a normal subset of $G$, \item $A$ acts on $B$ by conjugation, and \item the quotient group $G/B$ is isomorphic to $A$. Note that the order of the factors is significant.\end{inparaenum} The semidirect product is not unique, but depends on the choice of $\varphi$. If $\varphi_a$ is the identity map for all $a$, the semidirect product is reduced to the direct product.

\subsection{Short Exact Sequences}

The most general way to describe a group embedding is a short exact sequence, as we will now explain. An \textit{exact sequence} is a collection of groups and homomorphisms
\begin{equation}
G_1 \overset{\varphi_1}{\longrightarrow} G_2 \overset{\varphi_2}{\longrightarrow} G_3 \overset{\varphi_3}{\longrightarrow} \ldots \overset{\varphi_{n-1}}{\longrightarrow} G_n
\end{equation}
such that the image of each homomorphism is equal to the kernel of the following one, i.e.~$\mathrm{Im}(\varphi_{k}) = \mathrm{Ker}(\varphi_{k+1})$. A \textit{short exact sequence} is an exact sequence of the form
\begin{equation}
\bs{1} \overset{\varphi_0}{\longrightarrow} G_1 \overset{\varphi_1}{\longrightarrow} G_2 \overset{\varphi_2}{\longrightarrow} G_3 \overset{\varphi_3}{\longrightarrow} \bs{1},
\end{equation}
where $\bs{1}$ denotes the trivial group. A group homomorphism always maps the identity element onto the identity element, so $\mathrm{Im}(\varphi_{0})=\mathds{1}$. Because the sequence is exact, we have $\mathrm{Ker}(\varphi_{1})=\mathrm{Im}(\varphi_{0}) = \mathds{1}$, i.e.~$\varphi_{1}$ is injective. Since $\varphi_{3}$ maps everything to $\mathds{1}$, its kernel is $G_3$, and by the same argument we can conclude that $\mathrm{Im}(\varphi_{2}) = \mathrm{Ker}(\varphi_{3})=G_3$, i.e.~$\varphi_{2}$ is surjective.

\medskip

The isomorphism theorem states that for any homomorphism $\varphi: A \rightarrow B$,
\begin{equation}
\mathrm{Im}(\varphi) = A \big/ \mathrm{Ker}(\varphi).
\label{eq:isomorphism_theorem}
\end{equation}
It is easy to see why this holds: $\varphi$ is into $B$, but onto $\mathrm{Im}(\varphi)\subset B$, so $\varphi: A\rightarrow \mathrm{Im}(\varphi)$ is surjective. The kernel $\mathrm{Ker}(\varphi)\subset A$ is in general not trivial, so $\varphi$ is not injective. Dividing $A$ by the kernel (which is always a normal subset) identifies all elements in $\mathrm{Ker}(\varphi)$ with $\mathds{1}$, so $\varphi: A/\mathrm{Ker}(\varphi)\rightarrow \mathrm{Im}(\varphi)$ becomes injective. This concludes the heuristic proof of \ref{eq:isomorphism_theorem}.

\medskip

Applied to our case, we obtain
\begin{equation}
\mathrm{Im}(\varphi_2) = G_2 / \mathrm{Ker}(\varphi_2) \quad \leftrightarrow \quad G_3 = G_2 / \mathrm{Im}(\varphi_1)\quad \leftrightarrow \quad G_3 = G_2 / G_1.
\label{eq:definition_of_group_extension}
\end{equation}

For the last equivalence we have used the fact that $\varphi_1$ is injective and thus establishes an isomorphism between  $G_1$ and $\mathrm{Im}(\varphi_1)$. Again, since the kernel of a homomorphism is always normal, $G_1 \simeq \mathrm{Im}(\varphi_1)=\mathrm{Ker}(\varphi_2)$ is a normal subset of $G_2$.

\medskip

We rewrite \ref{eq:definition_of_group_extension} using more suggestive notation:

\begin{equation}
\bs{1} \rightarrow N  \overset{\varphi}{\rightarrow} G \overset{\psi}{\rightarrow} Q \rightarrow \bs{1} \qquad \Rightarrow \qquad N \vartriangleleft G \quad\text{and}\quad Q = G / N
\label{eq:definition_of_group_extension_2}
\end{equation}

\ref{eq:definition_of_group_extension_2} gives a description of $G$ in terms of a normal subgroup $N$ and the quotient group $G/N$ and we say that $G$ is an extension of $Q$ by $N$.

\clearpage
\newpage

\bibliography{mybibliography}

\bibliographystyle{./utphys}

\end{document}